\UseRawInputEncoding
\RequirePackage[mathlines]{lineno}
\documentclass[a4paper,11pt]{article}
\pdfoutput=1
\usepackage{jheppub}

\usepackage[T1]{fontenc}

\usepackage{threeparttable}
\usepackage{multirow}
\usepackage{subfigure}
\usepackage{float}
\usepackage{subfigure}
\usepackage{booktabs}
\let\oldequation\equation
\let\oldendequation\endequation
\renewenvironment{equation}
  {\linenomathNonumbers\oldequation}
  {\oldendequation\endlinenomath}
\usepackage{lineno}
 \usepackage{amsmath}
\usepackage{graphicx}
\usepackage{epstopdf}

\def \jpsi {J/\psi}

\def \ee   {e^+e^-}

\def \piz  {\pi^0}
\def \pip  {\pi^+}
\def \pim  {\pi^-}

\def \gevcc{\mbox{GeV/$c^2$}}

\newcommand{\pp}{\pi^+\pi^-}
\newcommand{\EE}{e^+e^-}
\newcommand{\psp}{\psi(3686)}

\begin{document}

\title{\boldmath Study of the $e^+e^- \to \pi^{+}\pi^{-}\omega$ process at center-of-mass energies between 4.0 and 4.6 GeV}

\collaborationImg{\includegraphics[height=30mm,angle=90]{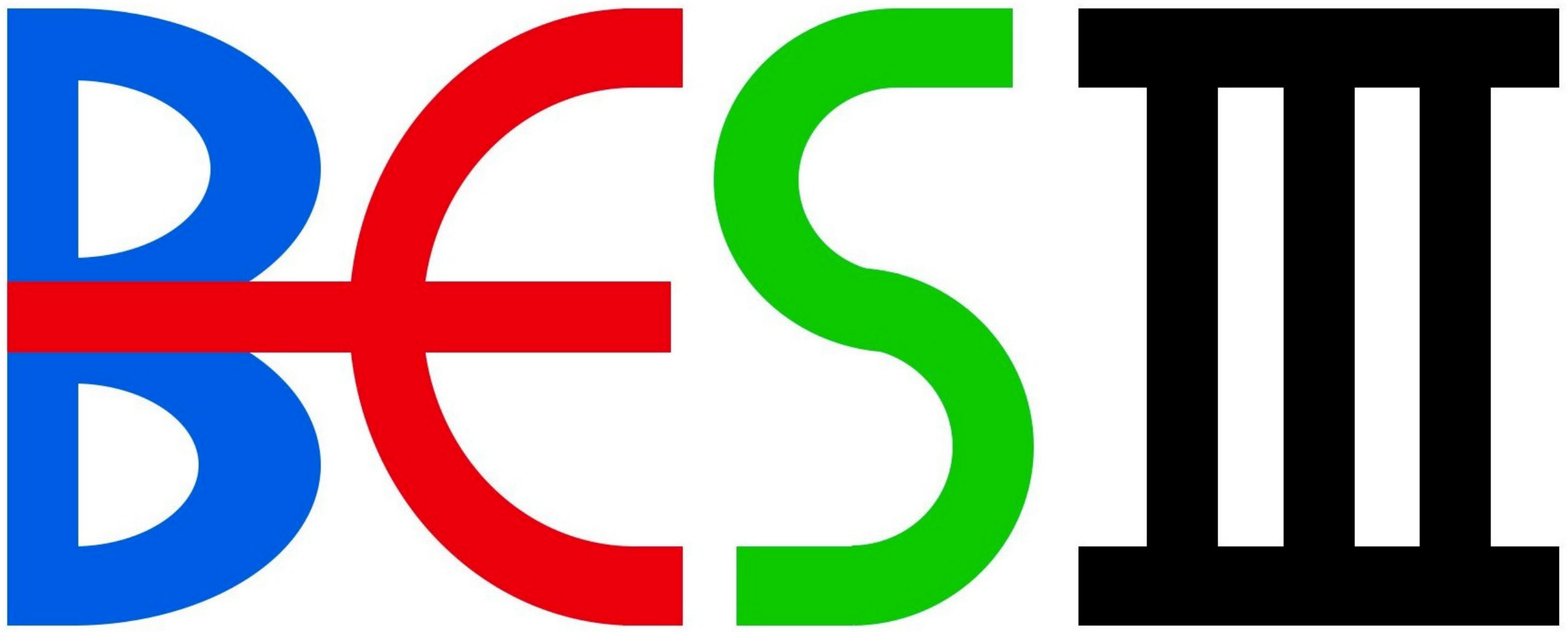}}
\collaboration{The BESIII collaboration}

\abstract{Using $15.6$ $\rm fb^{-1}$ of $e^+e^-$ collision data collected at
twenty-four center-of-mass energies from $4.0$ to $4.6$ GeV with the BESIII detector,
the helicity amplitudes of the process $e^+e^-\to \pi^{+}\pi^{-}\omega$ are analyzed for the first time.
Born cross section measurements of two-body intermediate resonance states with statistical
significance greater than 5$\sigma$ are presented, such as $f_{0}(500)$,
$f_{0}(980)$, $f_{2}(1270)$, $f_{0}(1370)$, $b_{1}(1235)^{\pm}$, and
$\rho(1450)^{\pm}$.
In addition, evidence of a resonance state in $e^+e^-\to \pi^+\pi^-\omega$
production is found.
The mass of this state obtained by line shape fitting
is about $4.2$~GeV/$c^2$, which is consistent with the production
of $\psi(4160)$ or $Y(4220)$.}

\maketitle
\flushbottom

\section{INTRODUCTION}
\label{sec:introduction}
\hspace{1.5em}
In recent years the study of charmonium-like($XYZ$) states has become a hot topic for both
experimental and theoretical physics due to their unexpected resonance parameters and
exotic decay patterns~\cite{pdg}.
Since $2003$, a series of charmonium-like states inconsistent with the quark model, such as the $X(3872)$~\cite{ref5},
$Y(4260)$~\cite{ref6} and $Z_c(3900)$~\cite{ref7,ref8}, have been observed.
In particular, the vector charmonium-like state $Y(4260)$ was observed by the BaBar experiment
in $\ee\to\gamma_{\rm ISR}\pp\jpsi$~\cite{ref6} and was confirmed by the CLEO and Belle
experiments~\cite{ref9,ref10}. In $2017$, the BESIII experiment performed a dedicated scan
of $\ee\to\pp\jpsi$ and observed two structures in this energy region. The
one with the mass $M = (4222.0\pm3.1\pm1.4)$ MeV/$c^2$~\cite{ref11} was regarded as
the previously observed $Y(4260)$, and renamed as $Y(4220)$. The $Y(4220)$ was then
confirmed in the Born cross section line shapes of $e^+e^-\to\omega\chi_{c0}$~\cite{ref12},
$\pp h_c$~\cite{ref13}, $\pp\psi(3686)$~\cite{ref14}, and $\pip D^0D^{*-}$~\cite{ref15} measured by the BESIII experiment. The other structure was identified with the $Y(4360)$, which was previously observed in
$\EE\to\gamma_{\rm ISR}\pp\psp$ by the BaBar experiment in $2007$~\cite{ref16}. Theoretically, many assignments, such as a tetraquark state~\cite{ref3,ref4,a1,a2,a3,a4,a5,a6},
a hybrid state~\cite{ref2,b1,b2,b3,b4}, a hadro-charmonium state~\cite{c1,c2,c3,c4}, a molecular
state~\cite{d1,d2,d3,d4}, a kinematic effect~\cite{e1,e2,e3,e4}, a baryonium state~\cite{f1}, etc., were proposed to explain the $Y$ state.

The traditional charmonium states, such as $\psi(4160)$ and $\psi(4040)$, were
observed in $\ee\to \rm hadrons$~\cite{dasp, bes2, rvue}
and $B^+\to K^+\mu^+\mu^-$~\cite{psi4160}.
However, their decays into light hadron final states have never been observed.
Many searches have been performed for these charmonium(-like) states
produced in $e^+e^-$ collisions and decaying to light hadron final states,
including $K_S^0K^\pm\pi^\mp\pi^0/\eta$~\cite{kkpipi0},
$K_S^0K^\pm\pi^\mp$~\cite{kkpi}, $2(p\bar{p})$~\cite{4p}, and $\pi^+\pi^-\pi^+\pi^-\pi^0$~\cite{4040}.
Only evidence for $\psi(4040)\to \pi^+\pi^-\pi^+\pi^-\pi^0$ has been reported.

In this paper, we measure the Born cross sections of $e^+e^-\to \pi^{+}\pi^{-} \omega$
at 24 center-of-mass (c.m.)~energies between 4.0 and 4.6 GeV, to search for the
charmonium(-like) states decaying into light hadron final sates.
Furthermore, we study intermediate states in the $e^+e^-\to \pi^+\pi^-\omega$ process via partial wave analysis (PWA).

\section{BESIII DETECTOR AND MONTE CARLO SIMULATION}
\label{sec:detector}
\hspace{1.5em}

The BESIII detector is a magnetic spectrometer~\cite{BESIII} located at
the Beijing Electron Positron Collider~(BEPCII) ~\cite{BEPCII}. The cylindrical
core of the BESIII detector consists of a helium-based multilayer drift
chamber (MDC), a plastic scintillator time-of-flight system (TOF), and
a CsI(Tl) electromagnetic calorimeter (EMC), which are all
enclosed in a superconducting solenoidal magnet providing a $1.0$~T
magnetic field\cite{detvis}. The solenoid is supported by an octagonal flux-return
yoke with resistive plate counter muon identifier modules interleaved
with steel. The acceptance of charged particles and photons is $93\%$
over $4\pi$ solid angle. The charged-particle momentum resolution at
$1~{\rm GeV}/c$ is $0.5\%$, and the specific ionization energy loss
(${\rm d}E/{\rm d}x$) resolution is $6\%$ for the electrons from
Bhabha scattering. The EMC measures photon energies with a resolution
of $2.5\%$ ($5\%$) at $1$~GeV in the barrel (end cap) region. The time
resolution of the TOF barrel part is $68$~ps, while that of the end cap
part is $110$~ps. The end cap TOF system was upgraded in $2015$ with
multi-gap resistive plate chamber technology, providing a time resolution of
$60$~ps~\cite{tof}; about 84\% of the data used here benefits from
this improvement.

This analysis uses data sets taken at twenty-four c.m.~energies ranging from 4.0 to 4.6 GeV.
For each data set, the c.m.~energy is calibrated by the di-muon process
$e^+e^- \to (\gamma_{\rm ISR,FSR})\mu^{+}\mu^{-}$~\cite{mumu},
where $\gamma_{\rm ISR,FSR}$ stands for possible initial state radiative (ISR) or final state radiative (FSR) photons.
The integrated luminosity ($\mathcal{L}_{\rm int}$) is determined
using large-angle Bhabha events~\cite{rlum}, and the total integrated luminosity of all data sets is $15.6$ fb$^{-1}$.

The BESIII detector is modeled with a Monte Carlo (MC) simulation using the software framework BOOST~\cite{boost},
based on GEANT4~\cite{geant4}, which includes the geometric and material description of the BESIII detector~\cite{geo1,geo2}, the
detector response, and digitization models, as well as the detector running conditions and performances.
Simulated MC samples generated by a phase space (PHSP) model with {\sc kkmc}~\cite{kkmc} are used for efficiency corrections in the PWA, and the TOY MC samples with detector simulation generated by ConExc~\cite{besevtgen} are used to determine detection efficiencies used for the Born cross-section determinations. The TOY MC events are generated based on helicity amplitude model with parameters fixed to the PWA results. The inclusive MC sample generated at
$\sqrt{s}=4.178$ GeV with {\sc kkmc}~\cite{kkmc} is used to study the potential backgrounds.

\section{EVENT SELECTION AND BACKGROUND ANALYSIS}
\label{sec:analysis}
\hspace{1.5em}
For $e^+e^-\to \pi^{+}\pi^{-}\omega$, $\omega\to\pi^{+}\pi^{-}\pi^0$, $\pi^0\to\gamma\gamma$,
the final state is characterized by four charged pion tracks and two photons.
For each charged track, the distance of closest approach to the interaction point is required to
be within $10$ cm in the beam direction and within $1$ cm in the plane perpendicular to the
beam direction. The track polar angle ($\theta$) must be
within the fiducial volume of the MDC, $i.e.$, $|\rm{\cos\theta}|<0.93$.
Particle identification~(PID) for charged tracks combines the d$E$/d$x$ and TOF information to form likelihoods
$\mathcal{L}(h)~(h=p,K,\pi)$ for each particle hypothesis.
Momentum-dependent PID is used to improve detection efficiency.
Charged tracks with momentum less than 0.9 GeV/$c$, are identified as pion candidates if their likelihoods satisfy ${\mathcal L}(\pi)>{\mathcal L}(K)$ and ${\mathcal L}(\pi)>{\mathcal L}(p)$. Those with momentum greater than $0.9$ GeV/$c$ are assigned as pion candidates with no PID requirement.

Isolated EMC showers are considered as photon candidates.
The deposited energy of each shower must be above $25$ MeV in the barrel region ($|\cos\theta|<0.80$) and $50$ MeV in the end cap region ($0.86<|\cos\theta|<0.92$).
Showers are required to occur within $700$~ns of the event start time to suppress noise.
Photon pairs with an invariant mass in the interval $0.11\sim 0.15$ $\gevcc$
are taken as $\pi^0$ candidates.

To reduce potential peaking backgrounds from $e^+e^-\to \gamma\omega$ with
$\gamma$ converting to $e^+e^-$,
the $\rm E_{EMC}/\it p$ of the pion candidate from non-$\omega$ decay is required to be less than $0.9$, where $\it p$ and $\rm E_{EMC}$ are momentum and EMC energy deposit associated with the track, respectively.
To suppress the backgrounds from $e^+e^-\to K_S^0\pip\pim\pi^0$ and $e^+e^-\to\chi_{c0}\omega$~\cite{xc0}, the invariant mass of all four $\pi^+\pi^-$ combinations are required to be outside the range of
($0.49$, $0.51$) and ($3.39$, $3.44$) GeV/$c^2$, respectively.
To further suppress the background and improve the mass resolution, we perform a five-constraint ($5$C) kinematic fit to the known initial four-momentum and $\piz$ mass~\cite{pdg}.
The $\chi^2_{\rm 5C}$ under the hypothesis of $e^+e^-\to \pi^+\pi^-\pi^+\pi^-\pi^0$ with $\pi^0\to \gamma\gamma$ is required to be less than $60$.
If more than one combination satisfies the above selection requirements,
only the one with the smallest $\chi^2_{\rm 5C}$ is kept.
To suppress background contribution from the final states with an additional photon, the $\chi^2_{\rm 5C}$ under the
$\pi^{+}\pi^{-}\pi^{+}\pi^{-}\pi^0$ hypothesis is required to be less than that under the $\pi^{+}\pi^{-}\pi^{+}\pi^{-}\pi^0\gamma$ hypothesis:
$\chi^2_{\rm 5C}(\pi^{+}\pi^{-}\pi^{+}\pi^{-}\pi^0) < \chi^2_{\rm 5C}(\pi^{+}\pi^{-}\pi^{+}\pi^{-}\pi^0\gamma)$.

\begin{figure}[htbp] \centering
	\setlength{\abovecaptionskip}{-1pt}
	\setlength{\belowcaptionskip}{10pt}
	\includegraphics[width=10.0cm]{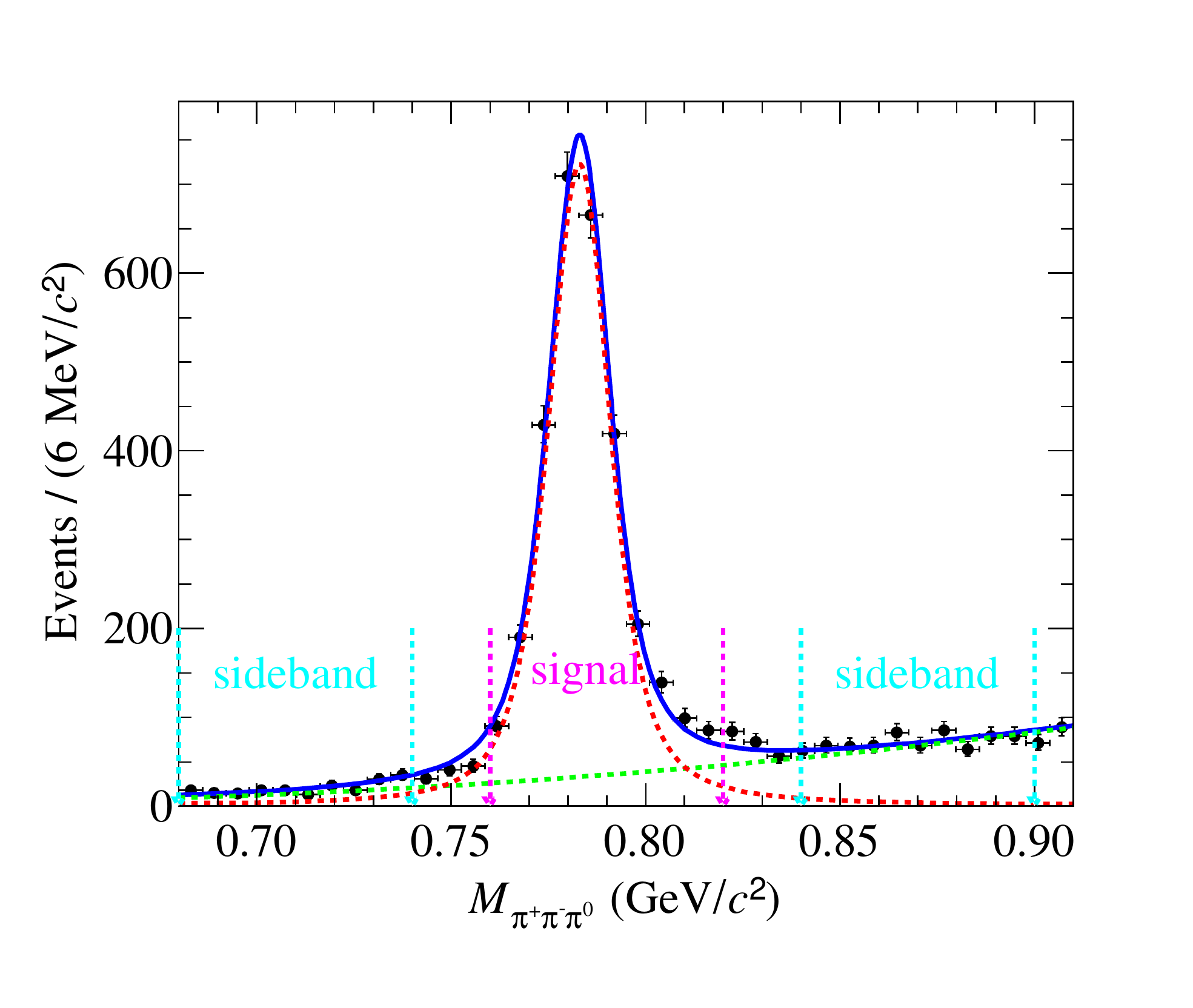}
	\caption{Distribution of $M_{\rm{\pip\pim\pi^0}}$ for events selected at $\sqrt{s}=4.1780$ GeV (black points with error bars).
The blue solid curve is the total fit result,
the red dashed curve is the fitted signal shape and the green dashed curve is the fitted background shape. The region between two dashed pink arrows is the $\omega$ signal region, while the regions between the pairs of neighboring dashed blue arrows are the $\omega$ sideband regions.}
	\label{fig:Mw}
\end{figure}

In each event, there are four $\pi^+\pi^-\pi^0$ combinations; the one with the invariant mass $M_{\rm \pip\pim\pi^0}$ closest to the known $\omega$ mass~\cite{pdg} is chosen as the $\omega$ candidate. This may distort the combinatoric background shape.
A study of an $e^+e^-\to\pi^+\pi^-\pi^+\pi^-\pi^0$ MC sample leads to a smooth distribution of
the invariant mass of combinatoric $\pip\pim\pi^0$
that can be described by a polynomial function.
A study based on the signal MC sample shows that the ratio of the yield of combinatoric $\pi^+\pi^-\pi^0$ background to the signal yield is $1.4$\% and results in a negligible difference of $0.1$\% on the fitted signal yield.
Figure~\ref{fig:Mw} shows the $M_{\rm \pip\pim\pi^0}$ distribution of the accepted
events from the data sample taken at $\sqrt{s}=4.178$ GeV.
To extract the number of signal events,
an unbinned extended maximum-likelihood fit is performed on the $M_{\rm \pip\pim\pi^0}$ distribution.
The signal shape is a MC-derived shape convolved with an additional Gaussian
smearing function, and the background shape is
a second-order Chebychev polynomial function.
The signal yields are listed in Table~\ref{tab:sec}.  Based on the $M_{\rm \pip\pim\pi^0}$ resolution from fitting,
the $\omega$ signal region is defined as $M_{\pi^+\pi^-\pi^0}\in (0.76, 0.82)$ GeV/$c^2$, while the $\omega$ sideband regions are defined as
the regions $M_{\pi^+\pi^-\pi^0}\in (0.68,0.74)$ GeV/$c^2$ and $M_{\pi^+\pi^-\pi^0}\in (0.84, 0.90)$ GeV/$c^2$.

\vspace{-0.0cm}
\vspace{-0.0cm}

\section{AMPLITUDE ANALYSIS}
\label{sec:amplitude}

\subsection{Kinematic variable and helicity angles}
\label{sec:amplitude}
\begin{figure}[htbp] \centering
	\setlength{\abovecaptionskip}{-1pt}
	\setlength{\belowcaptionskip}{10pt}
\includegraphics[width=6.0in,angle=0]{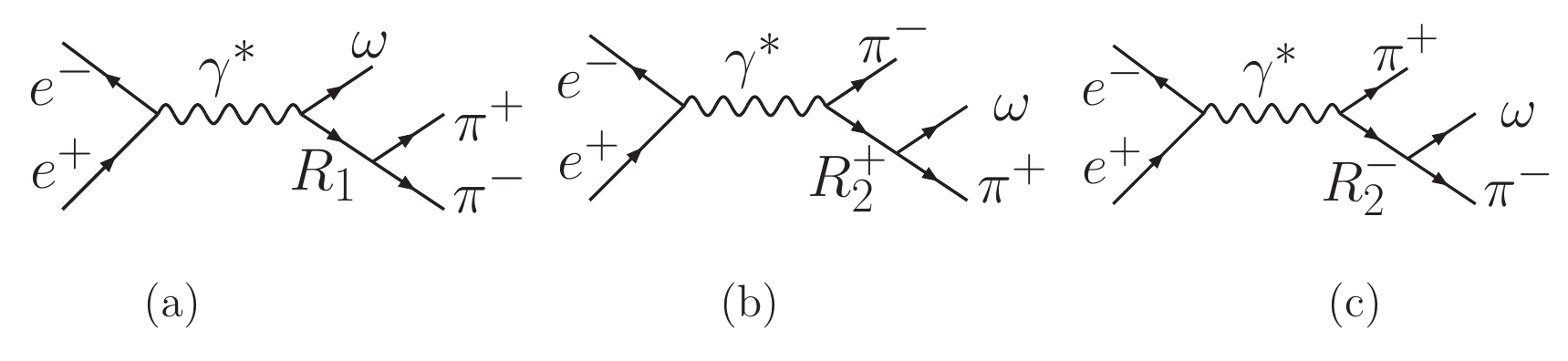}
\put(-280,90){\rotatebox{90}} \caption{The Feynman diagrams of quasi-two body decays in the process $e^{+} e^{-}\to\pi^{+}\pi^{-}\omega$ with different subprocesses: (a) $\ee\to R_1\omega, ~R_1\to\pip\pim$; (b) $\ee\to \pim R_2^+, ~R_2^+\to \pip\omega$;
and (c) $\ee\to \pip R_2^-, ~R_2^-\to \pi^-\omega$, where $R_1$ and $R_2^\pm$ denote intermediate states.
} \label{figfm}
\end{figure}

\hspace{1.5em}
The $\pip\pim\omega$ final sate is produced from the $\ee$ annihilation into a virtual photon,
followed by hadronization into the $\pip(p_1)\pim(p_2)\omega(p_3)$ final sate, where
$p_i$ ($i=$ $1$, $2$, $3$) denote particle momenta after the kinematic fit.
The $\pi^+\pi^-\omega$ final state may be produced non-resonantly,
or via an intermediate resonance and subsequent decay;
the possible resonance diagrams are shown in Fig.~\ref{figfm}.

\begin{figure}[h!]
\begin{center}
\includegraphics[width=4.2in,angle=0]{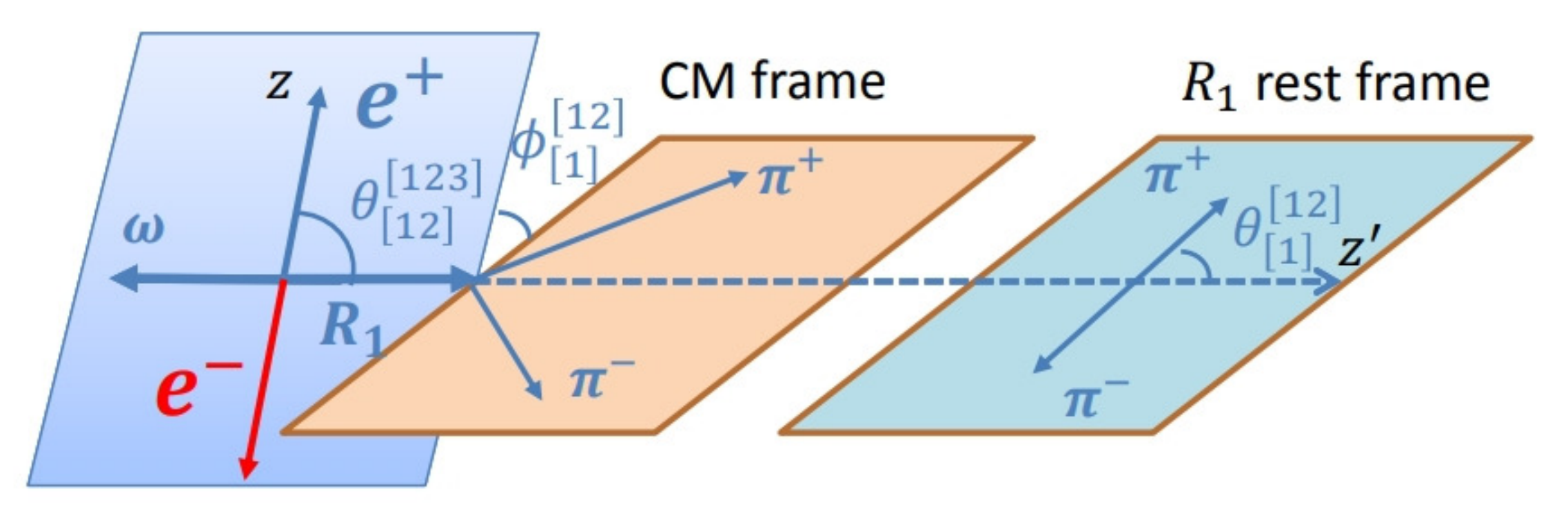}
\put(-280,90){\rotatebox{90}} \caption{Definitions of helicity rotation angles for the process $e^{+} e^{-} \to R_1\omega,~R_1\to  \pi^+\pi^-$.} \label{helicity1}
\end{center}
\end{figure}

The amplitudes for these diagrams are constructed
using the helicity formalism. Taking the first diagram in Fig.~\ref{figfm} as an example, one may define the helicity rotation angles as in Fig.~\ref{helicity1}. For resonance
$R_1$ the polar angle
($\theta_{[12]}^{[123]}$) is defined as the angle spanned between the $R_1$ momentum
and the positron beam direction, the azimuthal angle
($\phi_{[12]}^{[123]}$) is the angle between the $R_1$
production plane formed by the $R_1$ momentum and the $z$ axis and the plane formed by the $x$ and $z$ axes.
Here, $xyz$ denotes the laboratory coordinates.
The helicity amplitude for $\gamma^*\to R_1 (\lambda_R)\omega(\lambda_3) $
is denoted by $F^{\gamma^*}_{\lambda_R, \lambda_3}$ with specified helicity $\lambda_R$ and $\lambda_3$.
For the $R_1\to\pip\pim$ decay,
the azimuthal angle ($\phi_{[1]}^{[12]}$)
is defined as the angle between the $R_1$ production plane and its decay plane, formed by
the momenta of $\pip\pim$ from $R_1$. After boosting the two pion momenta to the $R_1$ rest frame, they are still
located in the same decay plane. The polar angle ($\theta_{[1]}^{[12]}$)
for $\pip$ is defined as the angle between
the $\pip$ and $R_1$ momenta in the $R_1$ rest frame.
The helicity amplitude of this decay is denoted by $F^{R_1}_{0,0}$.
Helicity angles for the processes (b) and (c) are defined analogously.
Table~\ref{tab1} summarizes the helicity angles and amplitudes for
the three processes.

\vspace{-0.0cm}
\vspace{-0.0cm}

\begin{table}[h!]
\begin{center}
\caption{Variable definitions for the helicity angles and helicity amplitudes of the sequential processes (a), (b) and (c) shown in Fig.~\ref{figfm}. The $\lambda_i$ denotes the helicity value for the corresponding particle, and $m$ denotes the spin $z$ projection of virtual photon ($\gamma^{*}$) in electron-positron annihilation. \label{tab1} }
\vspace{0.2cm}
\begin{tabular}{lll}
\hline\hline
Process & Helicity angle & Helicity amplitude\\\hline
$\ee\to\gamma^*(m)\to R_1(\lambda_R)\omega(\lambda_3)$ & $\theta_{[12]}^{[123]},\phi_{[12]}^{[123]}$ & $F^{\gamma^*}_{\lambda_R,\lambda_3}$\\
$R_1\to \pi^+\pi^-$ & $\theta_{[1]}^{[12]},\phi_{[1]}^{[12]}$ & $F^{R_1}_{0,0}$\\\hline
$\ee\to\gamma^*(m)\to R_2^+ (\lambda_{+})\pi^-$ & $\theta_{[13]}^{[123]},\phi_{[13]}^{[123]}$ & $F^{\gamma^*}_{\lambda_{+},0}$\\
$R_2^+\to\omega(\lambda_3^{'})\pi^+$ & $\theta_{[3]}^{[13]},\phi_{[3]}^{[13]}$ & $F^{R_2^+}_{\lambda_3^{'},0}$\\\hline
$\ee\to\gamma^*(m)\to R_2^- (\lambda_{-})\pi^+$ & $\theta_{[23]}^{[123]},\phi_{[23]}^{[123]}$ & $F^{\gamma^*}_{\lambda_{-},0}$\\
$ R_2^- \to\omega(\lambda_3^{''})\pi^-$ & $\theta_{[3]}^{[23]},\phi_{[3]}^{[23]}$ & $F^{ R_2^- }_{\lambda_3^{''},0}$\\\hline\hline
\end{tabular}
\end{center}
\end{table}

\subsection{Decay amplitude}
\label{sec:amplitude}
\hspace{1.5em}
The decay amplitude for the process (a) is
\begin{equation}
A_1(m,\lambda_3)=\sum_{\lambda_R} F^{\gamma^*}_{\lambda_R,\lambda_3}D^{1*}_{m,\lambda_R-\lambda_3}(\phi_{[12]}^{[123]},\theta_{[12]}^{[123]},0)BW(m_{12})F^{R_1}_{0,0}D^{J*}_{\lambda_R,0}(\phi_{[1]}^{[12]},\theta_{[1]}^{[12]},0),
\end{equation}
where $D^J_{m,\lambda}(\phi,\theta,0)$ is the Wigner $D$-function, $J$ is the spin quantum number of resonance $R_1$, and $BW$ denotes the Breit-Wigner function.

The decay amplitude for the process (b) is
\begin{eqnarray}
A_2(m,\lambda_3)&=&\sum_{\lambda_+,\lambda_3'} F^{\gamma^*}_{\lambda_+,0}D^{1*}_{m,\lambda_+}(\phi_{[13]}^{[123]},\theta_{[13]}^{[123]},0)
BW(m_{13})F^{R_2^+}_{\lambda_3',0}D^{J*}_{\lambda_+,\lambda_3'}(\phi_{[3]}^{[13]},\theta_{[3]}^{[13]},0)\nonumber\\
&\times&D^{1}_{\lambda_3',\lambda_3}(\phi_3',\theta_3',0),
\end{eqnarray}
where $J$ is the spin of $R_2^+$. Since the $\omega$ helicity defined in the $R_2^+$ helicity system is different from that defined in the process (a), one needs to perform a rotation by the angles ($\theta_3', \phi_3'$) to align the $\omega$ helicity to coincide with that in the process (a). This issue has been addressed in the analyses \cite{lhcb,belle} and derived in detail in Ref. \cite{pingrg}.

The decay amplitude for the process (c) reads
\begin{eqnarray}
A_3(m,\lambda_3)&=&\sum_{\lambda_-,\lambda_3''} F^{\gamma^*}_{\lambda_-,0}D^{1*}_{m,\lambda_-}(\phi_{[23]}^{[123]},\theta_{[23]}^{[123]},0)
BW(m_{23})F^{ R_2^- }_{\lambda_3'',0}D^{J*}_{\lambda_-,\lambda_3''}(\phi_{[3]}^{[23]},\theta_{[3]}^{[23]},0)\nonumber\\
&\times&D^{1}_{\lambda_3'',\lambda_3}(\phi_3'',\theta_3'',0),
\end{eqnarray}
where the Wigner $D^{1}_{\lambda_3'',\lambda_3}(\phi_3'',\theta_3'',0)$ function is used to align the $\omega$ helicity to coincide with that defined in the process (a).

\begin{figure}[h!]
\begin{center}
\includegraphics[width=3in,angle=0]{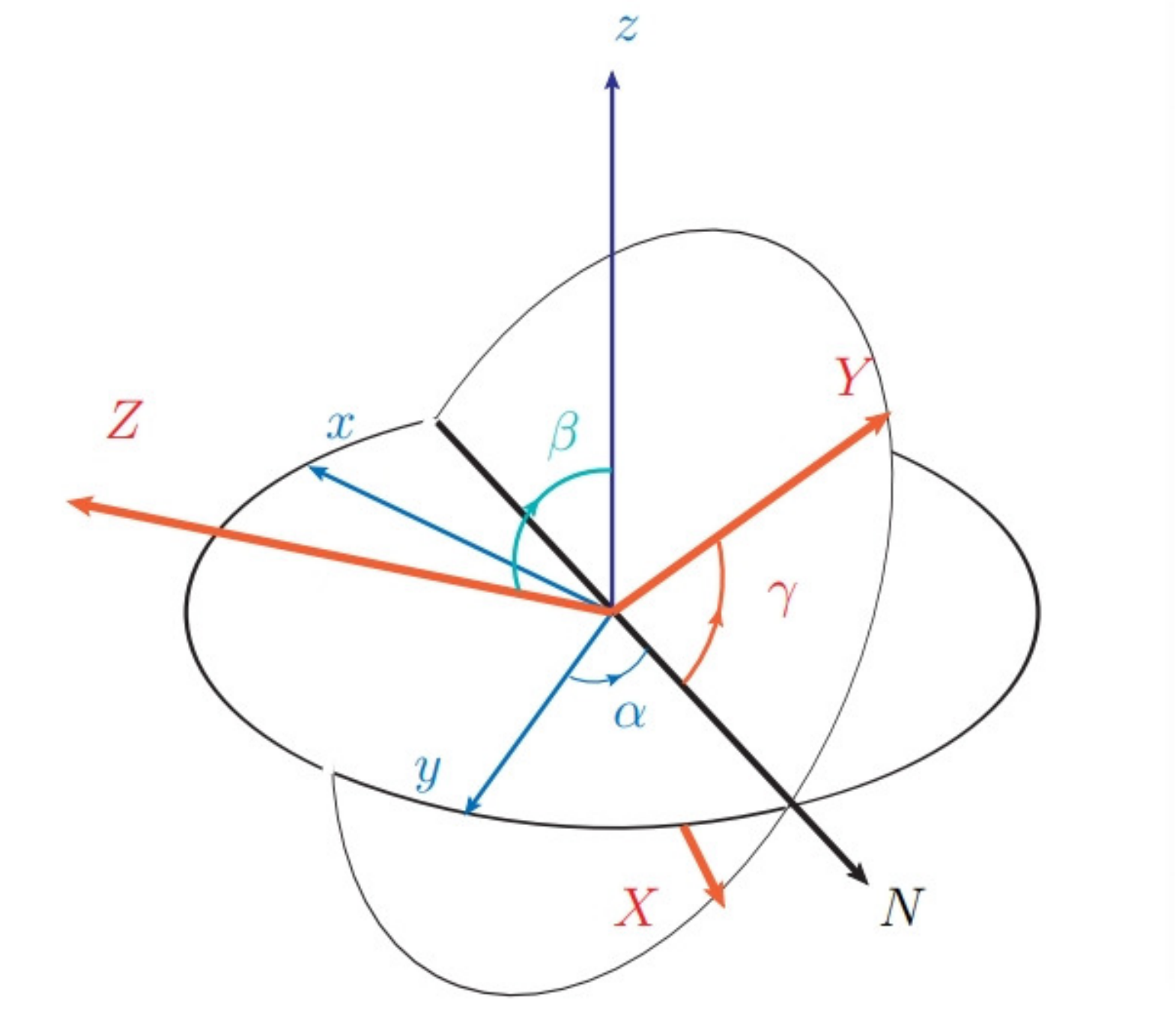}
\put(-280,90){\rotatebox{90}} \caption{The illustration of rotations to carry the $\omega$, $\pip$ and $\pim$ orientations from the rest
frame $xyz$ to the three-body helicity system $XYZ$ by the three Euler angles $\alpha$, $\beta$ and $\gamma$.} \label{helicity2}
\end{center}
\end{figure}

For the direct three-body process $\ee\to\pip\pim\omega$, the helicity amplitude is written as \cite{chung2}:
\begin{equation}
A_4(m,\lambda_3)=\sum_\mu{F_{\mu,\lambda_3}D^{1*}_{m,\mu}(\alpha,\beta,\gamma)},
\end{equation}
where $\mu$ is the $z$ component of the spin $J$ of the virtual photon in the helicity system,
and $m(\lambda_3)$ is the helicity value for $\gamma^*(\omega)$. Here,
$\alpha$, $\beta$, and $\gamma$ are the Euler angles as defined in \cite{chung2}
(see Fig.~\ref{helicity2}). $F_{\mu,\lambda_3}$ is the helicity amplitude;
parity conservation requires $F_{\pm,\lambda_3}=-F_{\pm,-\lambda_3}$
and $F_{0,\lambda_3}=F_{0,-\lambda_3}$. Parity conservation also requires
$F_{\pm,\lambda_3}(E_i) = -F_{\pm,-\lambda_3}(E_i)$ and $F_{0,\lambda_3}(E_i) = F_{0,-\lambda_3}(E_i)$, where $E_i (i=1, 2, 3)$ corresponds to the energy of the final state $\pip\pim\omega$.

One usually expands the helicity amplitudes in terms of the partial waves
for the two-body decay in the $LS$-coupling scheme~\cite{chung2}.
For a spin-$J$ particle decay $J\to s+\sigma$,
it follows
\begin{equation}\label{chung_forma}
F^J_{\lambda,\nu}  = \sum_{ls}\left ({2l+1\over 2J+1}\right )^{1/2}\langle l0S\delta|J\delta\rangle\langle s\lambda\sigma-\nu|S\delta\rangle g_{lS}r^l{B_l(r)\over B_l(r_0)},
\end{equation}
where $\lambda$ and $\nu$ are the helicities of two final-state particles $s$ and $\sigma$ with $\delta=\lambda-\nu$, and $g_{lS}$ is a coupling constant, $S$ is the total spin ${\bf S=s+\sigma}$, $l$ is the orbital angular momentum, $r=|{\bf r}|$, where ${\bf r}$ is the relative momentum between the two daughter particles in their mother rest frame, ${\bf r}_0$ corresponds to the value at the resonance's known mass. $B_l(r)$ is the Blatt-Weisskopf factor \cite{chung2}, which suppresses the contributions with higher angular momentum. The Blatt-Weisskopf factors up to $l$ = $4$  are
\begin{eqnarray}\label{}
B_0(r)/B_0(r_0)&=&1,\nonumber\\
B_1(r)/B_1(r_0)&=&\frac{\sqrt{1+(dr_0)^{2}}}{\sqrt{1+(dr)^{2}}},\nonumber\\
B_2(r)/B_2(r_0)&=&\frac{\sqrt{9+3(dr_0)^{2}+(dr_0)^{4}}}{\sqrt{9+3(dr)^{2}+(dr)^{4}}},\\
B_3(r)/B_3(r_0)&=&\frac{\sqrt{225+45(dr_0)^{2}+6(dr_0)^{4}+(dr_0)^{6}}}{\sqrt{225+45(dr)^{2}+6(dr)^{4}+(dr)^{6}}},\nonumber\\
B_4(r)/B_4(r_0)&=&\frac{\sqrt{11025+1575(dr_0)^{2}+135(dr_0)^{4}+10(dr_0)^{6}+(dr_0)^{8}}}{\sqrt{11025+1575(dr)^{2}+135(dr)^{4}+10(dr)^{6}+(dr)^{8}}}, \nonumber
\end{eqnarray}
where $d$ is a constant fixed to $3$ GeV$^{-1}$ for the meson final states~\cite{lhcb}.

The differential cross section is given by 	
\begin{equation}\label{eqq}
d\sigma=
\frac{1}{2}\sum_{m,\lambda_3}\Omega(\lambda_{3})\left|\sum_{j=1}^4 A_j(m,\lambda_3)\right|^2d\Phi,
\end{equation}
where $m=\pm1$ due to the polarization of the virtual photon being produced from $\ee$ annihilation, and $d\Phi$ is the element of standard three-body PHSP.
The $\Omega(\lambda_{3})=|{\boldsymbol \varepsilon}(\lambda_{3})\cdot ({\bf q_{1}} \times {\bf q_{2}})|^2$ is the $\omega$ decay matrix element into the $\pi^+\pi^-\pi^0$ final states,
where ${\boldsymbol \varepsilon}$ is the $\omega$ polarization vector, and $\bf q_{1}(q_{2})$ is the
momentum vector for $\pi^{+}(\pi^{-})$ from the $\omega$ decay. Here we factor out the $BW$ function describing
the $\omega$ line shape into the MC integration when applying the amplitude analysis to the data events.

\subsection{Simultaneous fit}
\label{sec:amplitude}
\hspace{1.5em}
The relative magnitudes and phases of the coupling constants are determined by an unbinned maximum likelihood fit. The joint probability density function (PDF) for the events observed in the data sample is defined as
\begin{equation}
\mathcal{L}=\prod_{i=1}^N P_i(p_1, p_2, p_3, p_4, p_5),
\end{equation}
where $p_i$ ($i$ = $1$, $2$, \dots, $5$) denotes the four-vector momenta of the final state particles, and $P_i$ is a probability to produce the $i$-th event. The normalized $P_i$ is calculated from the differential cross section
\begin{equation}
P_i={(d\sigma /d\Phi)_i \over \sigma_{\rm MC}},
\end{equation}
where $\sigma_{\rm MC}$ is the normalization factor which is calculated with a large MC sample as
\begin{equation}
\sigma_{\rm MC}\approx{1\over N_{\rm MC}}\sum_{i=1}^{N_{\rm MC}}\left({d\sigma\over d\Phi}\right)_i,
\end{equation}
where $N_{\rm MC}$ is the number of events retained with the same selection criteria as for data sample.

For technical reasons, rather than maximizing $\mathcal{L}$, $S=-\ln\mathcal{L}$ is minimized using the package MINUIT~\cite{minuit}. To subtract the contribution of background, the $\ln\mathcal{L}$ function is replaced with
\begin{equation}
\ln\mathcal{L} = \ln\mathcal{L}_\textrm{data}-\ln\mathcal{L}_\textrm{bkg},
\end{equation}
where $\mathcal{L}_\textrm{data}$ and $\mathcal{L}_\textrm{bkg}$ are the joint PDFs for data and background, respectively.
The background events are obtained from the $\omega$ sideband regions mentioned in Section 3.

A simultaneous fit is performed to data sets collected at different c.m.~energies. The common
parameters for different data samples in this fit are the masses,
widths, and $\rm Flatt\acute{e}$ parameters for the resonances. The total function is taken as the sum of individual ones, $i.e.$,
\begin{equation}
S^{'}= -\sum_{j=1}^{M}\ln{\mathcal{L_\textrm{j}}} .
\end{equation}

The signal yield for the $i$-th resonance, $N_i$, can be estimated by scaling its cross section ratio $R_i$ to the number of net events
\begin{equation}\label{yieldsFormula}
N_i=R_i(N_\textrm{obs}-N_\textrm{bkg}),\textrm{~with~}R_i={\sigma_i\over \sigma_\textrm{tot}},
\end{equation}
where $\sigma_i$ is the cross section for the $i$-th resonance as defined in Eq.(\ref{eqq}),
$\sigma_\textrm{tot}$ is the total cross section, and $N_\textrm{obs}$ and $N_\textrm{bkg}$
are the numbers of observed events and background events, respectively.
In the simultaneous fit, the background events are taken from the $\omega$ sideband regions, and
the number $N_{\rm bkg}$ is estimated with the background PDF with the $\omega$ signal region (see Fig.~\ref{fig:Mw}).

The statistical uncertainty, $\Delta N_i$, associated with the signal yield $N_i$, is estimated according to the error propagation formula using the covariance matrix, $V$, obtained in the simultaneous fit, {\it i.e.}
\begin{equation}\label{staterr}
\Delta N_{i}^{2} = \sum_{m=1}^{N_\textrm{pars}}\sum_{n=1}^{N_\textrm{pars}}\left({\partial N_i\over \partial X_m}{\partial N_i\over \partial X_n}\right)_{\bf{X}={\bf \mu}}V_{mn}({\bf X}),
\end{equation}
where ${\bf X}$ is a vector containing parameters, and ${\bf \mu}$ contains the fitted values for
all parameters. The sum runs over all $N_\textrm{pars}$ parameters.

\subsection{Intermediate states in $\pp\omega$ final state}
\label{sec:amplitude}
\hspace{1.5em}
In the $\pip\pim$ and $\omega\pi^\pm$ mass spectrum, the $f_{0}(500)$, $f_{0}(980)$, $f_{2}(1270)$, $f_{0}(1370)$,
$b_{1}(1235)^{\pm}$, $\rho(1450)^\pm$, and $\rho(1570)^\pm$ resonances are included in the amplitude model. The $f_0(980)$ line shape is parameterized by the $\rm Flatt\acute{e}$ formula:
\begin{equation}\label{f0flatte}
BW_1(s)={1\over s-M^2+i(g_1\rho_{\pi\pi}(s)+g_2\rho_{K\bar K}(s))},
\end{equation}
where $\rho(s)=2k/\sqrt s$ and $k$ is the momentum of the $\pi$ or $K$ in the resonance rest frame,
$g_1$ and $g_2/g_1$ are fixed to the measured values $(0.138 \pm 0.010)$ $\rm GeV^2 $ and
$4.45 \pm 0.25$ ~\cite{besiia,besiib}, respectively. $M$ is the mass of $f_{0}(980)$ taken from the PDG~\cite{pdg}.

For the $BW_2$ function of a wide resonance, {\it e.g.}, $f_{0}(500)$, there are many parametrizations for
the energy-dependent width~\cite{besiia,besiib}, and we take the one used by the E791 Collaboration in the nominal fit,
\begin{equation}
BW_2(s)={1\over s-m_0^2+i\sqrt{s}\Gamma},
\;{\rm with } \; \Gamma=\sqrt{1-{4m_\pi^2\over s}}\Gamma_0,
\end{equation}
where $m_0$ is the nominal mass of the resonance, and $\Gamma_0$ is its width. For other resonances, such as $b_{1}(1235)^{\pm}$, $f_{0}(1370)$,  $f_{2}(1270)$, $\rho(1450)^\pm$, $\rho(1570)^\pm$,
their line shapes are described with the $BW_3$ function,
\begin{equation}
BW_3(s)={1\over s-m_0^2+i\sqrt{s}\Gamma},
\end{equation}
where the widths are fixed to the individual PDG values~\cite{pdg}.

Based on the signal events in the $\pip\pim\piz$ mass spectrum,
we select twelve c.m.~energy points with relatively large statistics.
We divide these selected points into two groups.
Group A includes the data sets taken at
$\sqrt s$ = $4.0076$, $4.1780$, $4.1890$, $4.1990$, $4.2093$, and $4.2188$ GeV,
and group B includes $\sqrt s$ = $4.2263$, $4.2358$, $4.2439$, $4.2580$, $4.2668$, and $4.4156$ GeV. To check
the significance of each resonance and determine the nominal solution, a simultaneous fit
is performed to the data from a given group.
In each group, the cross sections of these intermediate states are regarded to be energy-dependent, so the parameters responsible for the virtual photon $\gamma^*$ coupling to a given state are allowed to vary in the fit for various energy points, while the coupling constant parameters for the subsequent decay are taken as the common parameters for all energies. The conjugate modes share the same coupling constants.
The masses,
widths or $\rm Flatt\acute{e}$ parameters for the resonances of $f_{0}(500)$, $f_{0}(980)$, $f_{2}(1270)$, $f_{0}(1370)$, $\rho(1450)^\pm$, and $\rho(1570)^\pm$ are fixed to the measured values from PDG~\cite{pdg}, as given in Table~\ref{sigtable}. The mass and width of $b_1(1235)^\pm$ are floated due to large uncertainties.
Then its nominal solution is fixed as the fitted result.

The significance of each intermediate state is estimated by the changes of $-2\ln \mathcal L$ and the number of degrees of freedom (NDF) after removing it from the simultaneous fit.
We take the intermediate states with statistical significances greater than 5$\sigma$ in two groups as
the nominal solution, including $f_{0}(500)$, $f_{0}(980)$, $f_{2}(1270)$, $f_{0}(1370)$, $b_{1}(1235)^{\pm}$, and $\rho(1450)^\pm$, as shown in Fig.~\ref{fig:Agroup}.
It is found that the contributions from $f_{0}(500)$ and $b_{1}(1235)^\pm$ are the most significant,
as shown in the $M_{\pi^+\pi^-}$ and $M_{\omega\pi^{\pm}}$ spectra, respectively. The statistical significances for various intermediate resonances are shown in Table~\ref{sigtable}.

\begin{table}[htbp]
\setlength{\belowcaptionskip}{10pt}
\caption{Masses, widths and statistical significances for various intermediate resonances in $e^+e^-\to \pi^+\pi^-\omega$.}\label{sigtable}
\centering{
\begin{tabular}{ccccc}
\hline\hline
Resonance & Mass (MeV/$c^2$) & Width (MeV) & Group A & Group B\\\hline
$f_0(500)$         & 507 (400$\sim$550)          & 475 (400$\sim$700)             &27.8$\sigma$    &22.8$\sigma$\\
$f_0(980)$       &$990\pm20$   &$-$    &10.9$\sigma$    &6.4$\sigma$\\
$f_0(1370)$      &1350$\pm$150                 &200$\pm$50                      &6.2$\sigma$    &3.4$\sigma$\\
$f_2(1270)$      &$1275.5\pm0.8$               &$186.7\pm2.2$                   &9.3$\sigma$    &5.4$\sigma$\\
$b_{1}(1235)^{\pm}$        &$1179.0\pm9.0$              &$255.8\pm16.4$                    &31.8$\sigma$    &25.7$\sigma$\\
$\rho(1450)^\pm$ & 1465.0$\pm$25               &400$\pm$60                      &4.7$\sigma$     &6.9$\sigma$\\
$\rho(1570)^\pm$ &1570$\pm$70                  &144$\pm$90                      &4.3$\sigma$     &2.4$\sigma$\\
$\pi^+\pi^-\omega$ & $-$ & $-$ & 6.5$\sigma$ & 3.0$\sigma$ \\
\hline\hline
\end{tabular}
}
\end{table}

\subsection{Fit results}
\label{sec:amplitude}
\begin{figure}[!h]
\setlength{\abovecaptionskip}{-0.5cm}
\hspace*{-29pt}
\includegraphics[width=0.35\textwidth,trim={2cm 8cm 3.5cm 1cm},clip]{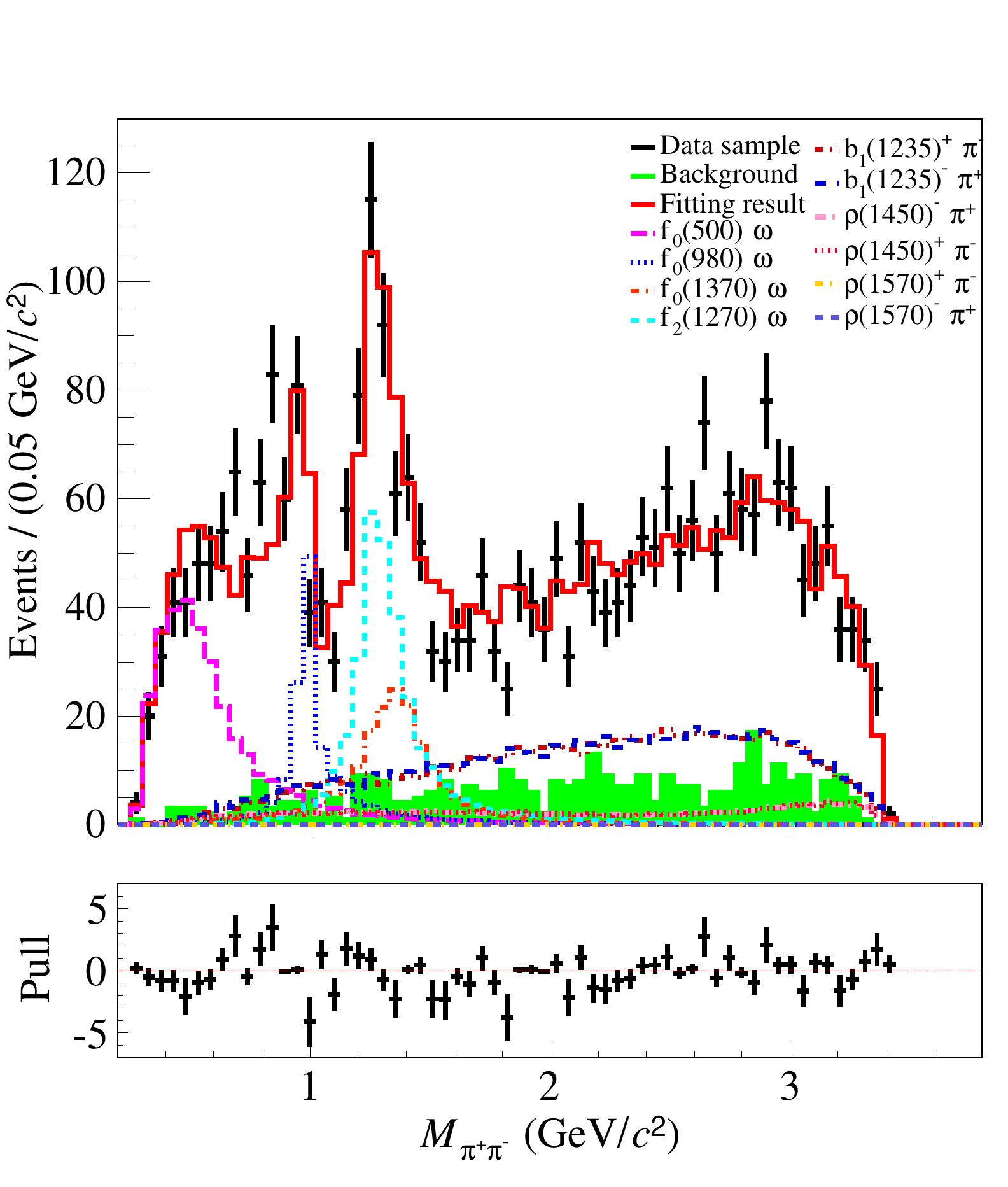}
\includegraphics[width=0.35\textwidth,trim={2cm 8cm 3.5cm 1cm},clip]{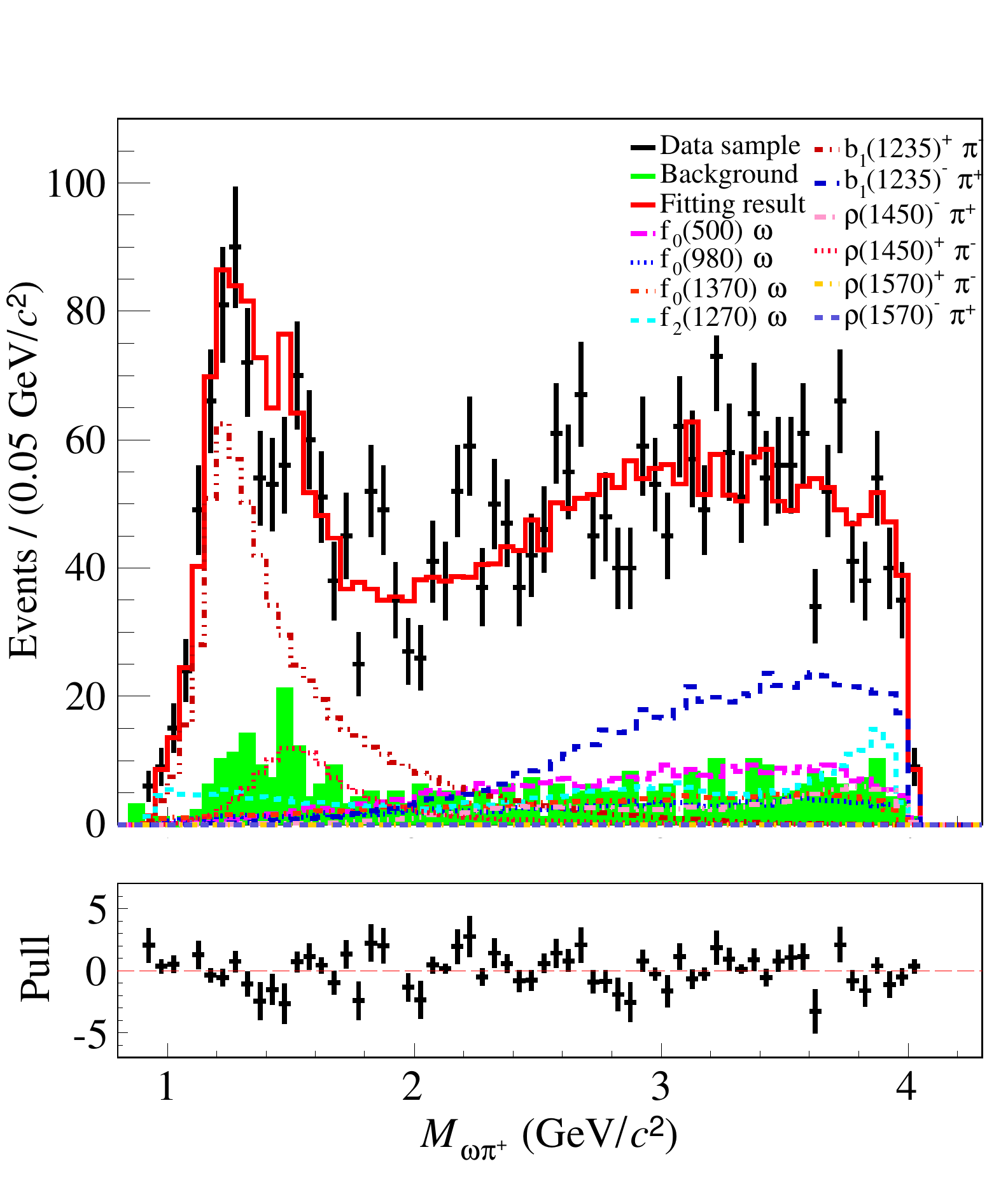}
\includegraphics[width=0.35\textwidth,trim={2cm 8cm 3.5cm 1cm},clip]{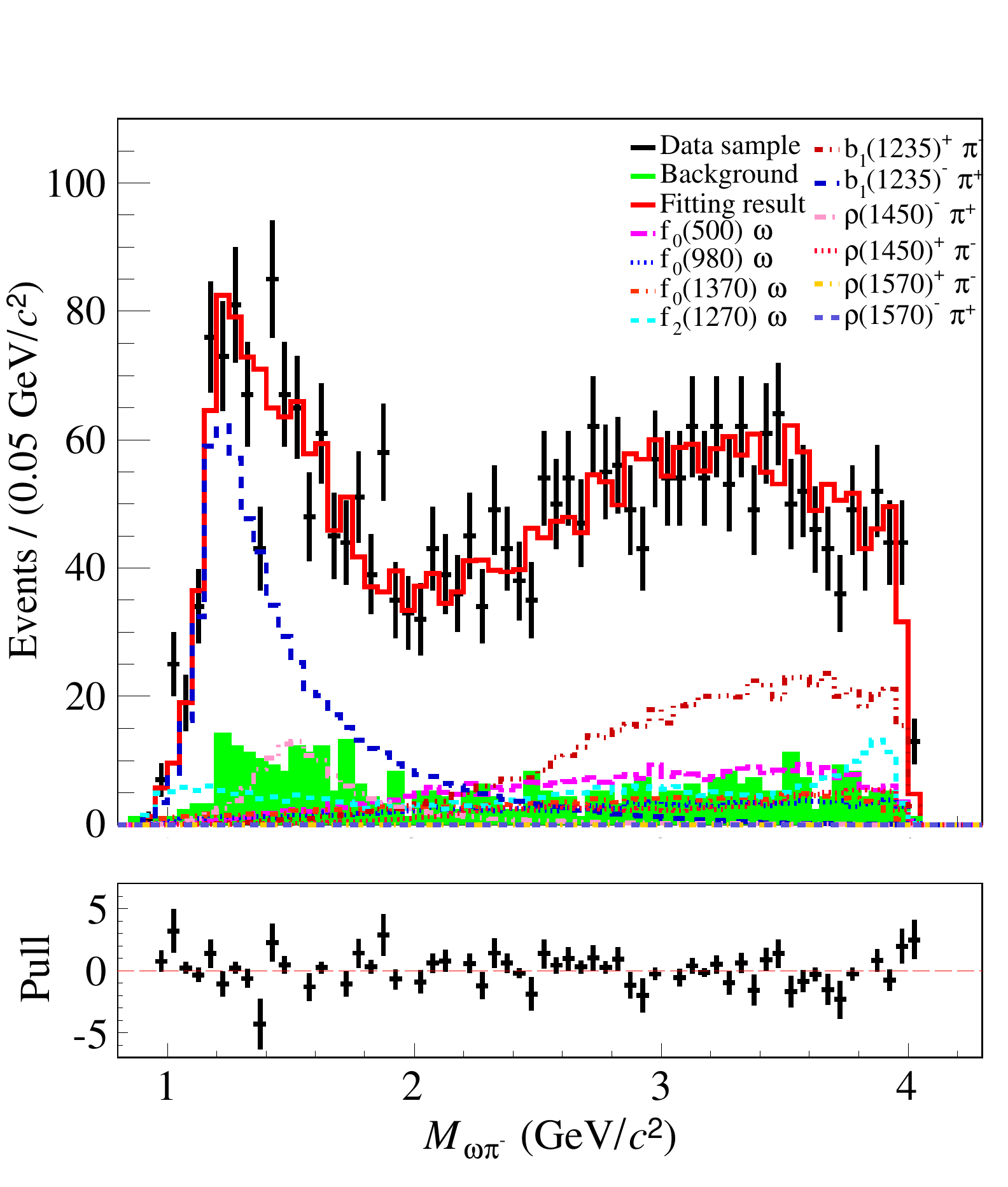}
\caption{ Projections of the PWA solution on the mass spectra $ M_{\pi\pi}$, $ M_{\omega\pi^{+}}$ and $ M_{\omega\pi^{-}}$
for the data sample collected at $\sqrt{s}$ = $4.1780$ GeV. Points with error bars are data, the red histogram shows the final PWA fit results,
and shaded histograms are the background estimated from the $\omega$ mass sideband regions.
Other line shapes marked with different colors represent the fitted line shapes of different intermediate resonance states.
The pull distribution of the fit result is shown at the bottom of each plot. }
\label{fig:Agroup}
\end{figure}

\begin{figure}[!h]
\setlength{\abovecaptionskip}{-0.5cm}
\hspace*{-29pt}
\includegraphics[width=0.35\textwidth,trim={2cm 8cm 3.5cm 1cm},clip]{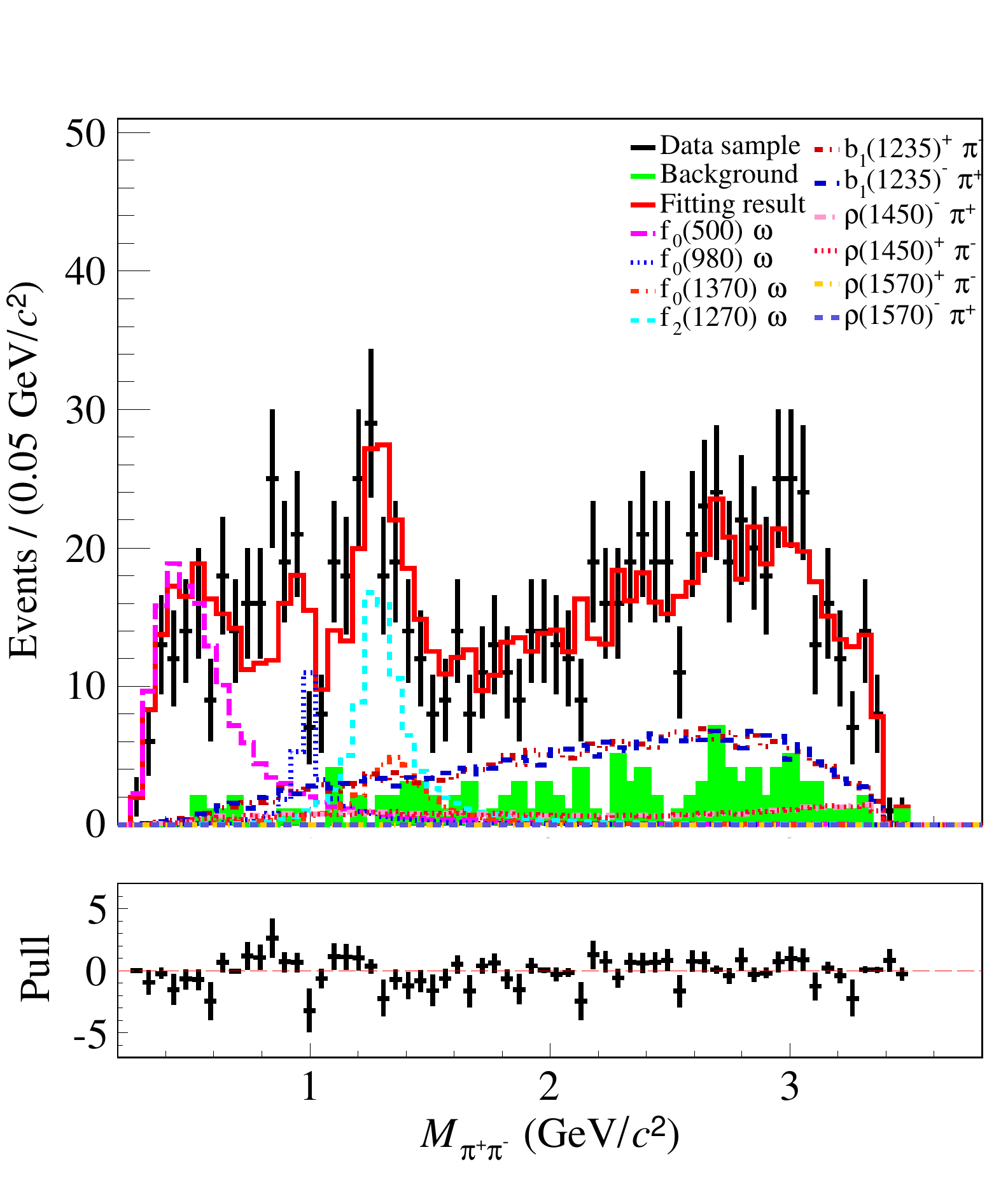}
\includegraphics[width=0.35\textwidth,trim={2cm 8cm 3.5cm 1cm},clip]{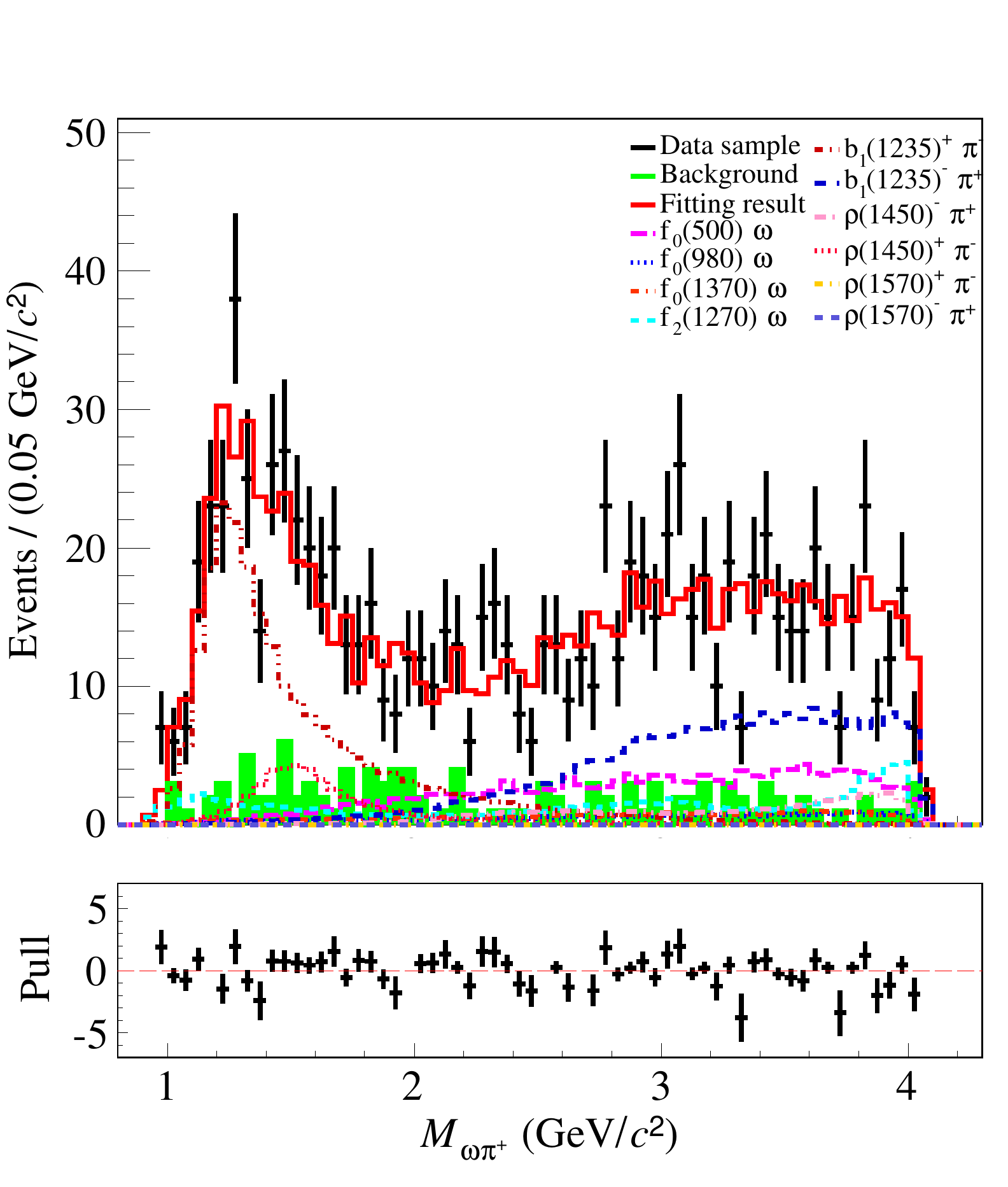}
\includegraphics[width=0.35\textwidth,trim={2cm 8cm 3.5cm 1cm},clip]{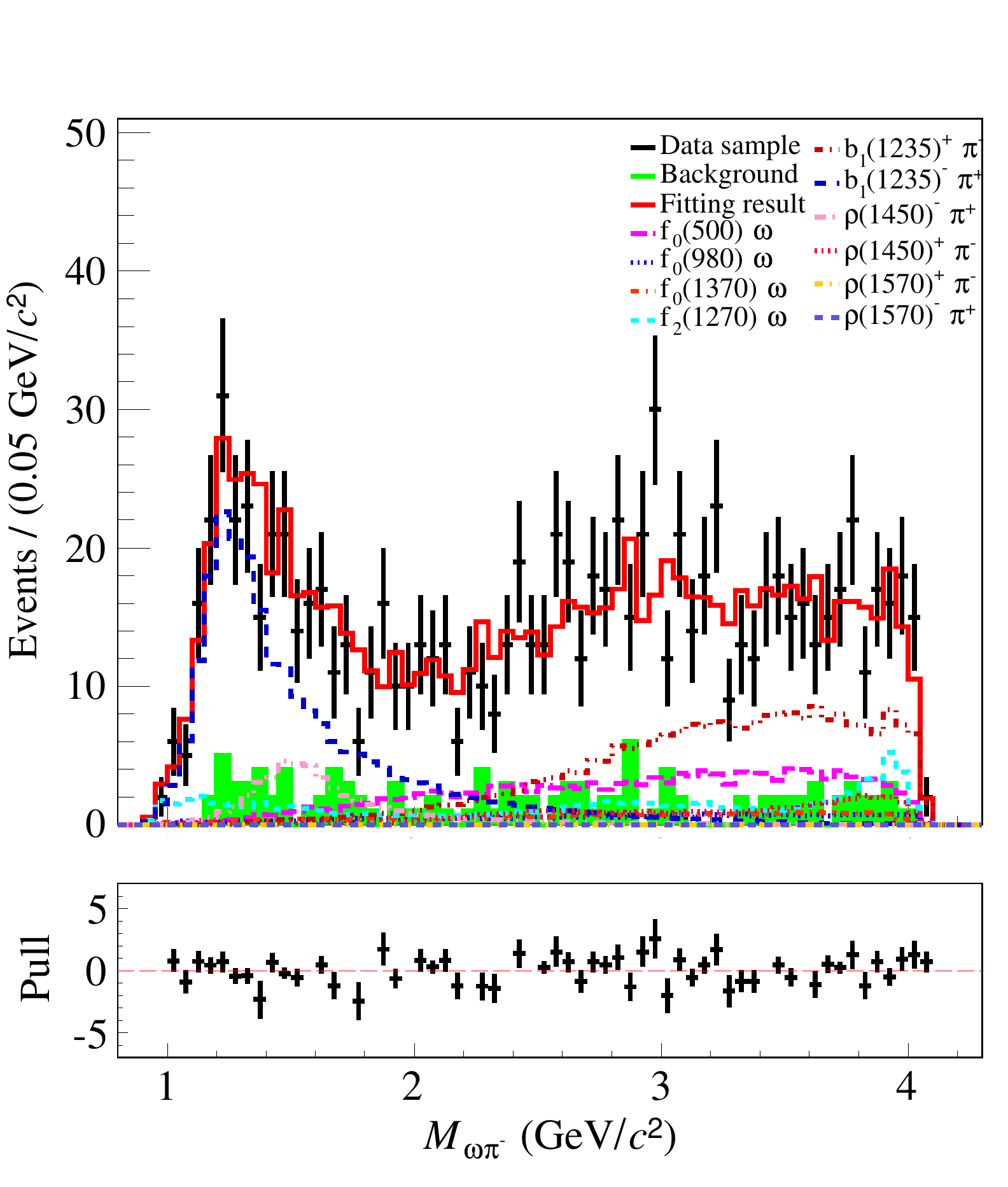}
\caption{ Projections of the PWA solution on the mass spectra $ M_{\pi\pi}$, $ M_{\omega\pi^{+}}$ and $ M_{\omega\pi^{-}}$ for the data sample collected at $\sqrt{s}$ = $4.2263$ GeV. Points with error bars are data, the red histogram shows the final PWA fit results, and shaded histograms are the background estimated from the $\omega$ mass sideband regions. Other line shapes marked with different colors represent the fitted line shapes of different intermediate resonance states. The pull distribution of the fit result is shown at the bottom of each plot. }
\label{fig:Bgroup}
\end{figure}

\hspace{1.5em}
For the simultaneous fit, the ratios and the signal yields of various intermediate states are obtained according to Eq.~(\ref{yieldsFormula}), as shown in Tables~\ref{tabratio} and~\ref{tabyield}. And their statistical uncertainties are determined based on Eq.~(\ref{staterr}), in which the correlation among parameters is included. With the intermediate states in the nominal solution, we perform the simultaneous fit to the data samples for groups A and B.  Taking the two data samples from $\sqrt s$ = $4.1780$ and $4.2263$~GeV with large integrated luminosity as examples,
Figs.~\ref{fig:Agroup} and ~\ref{fig:Bgroup} show the fit results for groups A and B, respectively.

\section{BORN CROSS SECTION}
\subsection{ISR correction factor}
\label{sec:born}
\hspace{1.5em}
In $\ee$ collision experiments, the observed cross section, $\sigma_{\rm obs}(s)$, at a c.m.~energy point $\sqrt{s}$, is related to the corresponding Born cross section, $\sigma_{0}(s)$, by the ISR factor
\begin{equation}\label{isr1}
1+\delta={\sigma_{\rm obs}(s)\over \sigma_0(s)},
\end{equation}
with
\begin{equation}\label{isr2}
\sigma_{\rm obs}(s)=\int_{M_{\rm th}}^{\sqrt{s}}W(s, x){\sigma_0[s(1-x)]\over{\vert 1-\Pi(\sqrt{s})\vert^2}}dx,
\end{equation}
where $\Pi(\sqrt{s})$ is the vacuum polarization (VP) function.
The $M_{\rm th}$ corresponds to the $\pip\pim\omega$ mass threshold, and $x$ is the effective fraction of the beam energy carried by photons emitted from the initial state, $x=\frac{2E_\gamma}{\sqrt{s}}$, and $E_\gamma$ is the
energy of the ISR photons.
The initial state radiative function, $W(s, x)$, which uses the QED
calculation up to next to leading order in Ref~\cite{wx,alpha,wsx},
\begin{equation}\label{isr3}
\begin{aligned}
W(s, x)&=\Delta\beta x^{\beta-1}-\frac{\beta}{2}(2-x)+\frac{\beta^2}{8}\Big\{(2-x)[3\ln(1-x) \\
&\phantom{=\;\;} -4\ln x]-4\frac{\ln(1-x)}{x}-6+x \Big\} ,
\end{aligned}
\end{equation}
where
\begin{equation}\label{}
\begin{split}
L&=2\ln{\frac{\sqrt{s}}{m_e}}, \\
\Delta&=1+\frac{\alpha}{\pi}\left(\frac{3}{2}L+\frac{1}{3}\pi^2-2\right)+\left(\frac{\alpha}{\pi}\right)^2\delta_{2}, \\
\delta_{2}&=\left(\frac{9}{8}-2\xi_2\right)L^2-\left(\frac{45}{16}-\frac{11}{2}\xi_2-3\xi_3\right)L-\frac{6}{5}\xi_2^2-\frac{9}{2}\xi_3-6\xi_2\ln2+\frac{3}{8}\xi_2+\frac{57}{12},\\
\beta&=\frac{2\alpha}{\pi}(L-1),~~\xi_2=1.64493407,~~\xi_3=1.2020569 ,
\end{split}
\end{equation}
and we use the calculated results including the leptonic and hadronic parts both in the space-like and time-like region~\cite{isr11, isr12, isr13, isr14, isr15}.

We use the generator model ConExc~\cite{besevtgen} to produce signal MC events and then iterate the Born cross-section measurement, in which the radiative function takes the result of high-order QED calculation up to the $\alpha^2$ accuracy~\cite{alpha}.
The Born cross sections from the $\pi^+\pi^-\omega$ mass threshold to 4.6 GeV are used to calculate the ISR factor. The Born cross sections $\sigma_{0}(s)$ in the c.m.~energy ranges of below $3.0$ GeV and $(4.0,4.6)$ GeV are taken from the measurements in Ref.~\cite{babarXS} and this work, respectively.  In the c.m.~energy interval of $(3.0,4.0)$ GeV, however, the Born cross section of $e^+e^-\to$ continuum light hadrons is described by a polynomial, and the Born cross sections for $J/\psi$ and $\psi(3686)$ are described by the function
\begin{equation} \label{eq:resonance}
\sigma(\sqrt{s}) = \frac{2J+1}{(2S_{1}+1)(2S_{2}+1)} \frac{4\pi}{k^{2}}\left[\frac{\Gamma^{2}/4}{(\sqrt{s}-\sqrt{s}_{0})^{2}+\Gamma^{2}/4}\right] B_{\rm in} B_{\rm out},
\end{equation}
where $\sqrt{s}$ is the c.m.~energy, $J=1$ is the spin of the resonance, and the numbers of polarization states of the two incident particles are $2S_{1}+1=2$ and $2S_{2}+1=2$, respectively. The maximum momentum of the final-state channel is denoted as $k$, $\sqrt{s}_{0}$ is the c.m.~energy at the resonance, and $\Gamma$ is the width of the resonance. The branching fractions of the resonance decays into the initial-state and final-state channels are denoted as $B_{\rm in}$ and $B_{\rm out}$, respectively.
The cross sections are smoothed by a fit to seven Gaussian functions in various energy intervals. 
Since the detection efficiency is affected by the radiative correction, an iteration over the cross section is done until the latest two results become stable; specifically, when the updated Born cross sections change by less than the statistical uncertainty.
The ISR correction factor for each c.m.~energy
point is given in Table~\ref{tab:sec}.

\subsection{Born cross section of $\ee \to \pip\pim\omega$}
\label{sec:born}
\hspace{1.5em}
The Born cross section at each c.m.~energy is calculated by
\begin{equation} \label{eq:sigmaobs}
\sigma^{\rm Born} = \frac{N^{\rm sig}}{\mathcal{L}_{\rm int}\cdot\epsilon \cdot (1+\delta^{\gamma})\cdot \frac{1}{\vert 1-\Pi\vert^2}\cdot Br(\omega\to\pi^+\pi^-\pi^0)\cdot Br(\pi^0\to\gamma\gamma)} ,
\end{equation}
where $N^{\rm sig}$ is the number of observed signal events, $(1+\delta^{\gamma})$ and $\frac{1}{\vert 1-\Pi\vert^2}$ are the ISR correction and VP corrections, respectively. The factors $Br(\omega\to\pi^+\pi^-\pi^0)$ and $Br(\pi^0\to\gamma\gamma)$ are the branching fractions of $\omega \to \pi^+\pi^-\pi^0$ and $\pi^0\to\gamma\gamma$ from the PDG~\cite{pdg}. We use $\epsilon$ to denote the detection efficiency determined by the TOY MC sample with detector simulation of helicity amplitude model.
The numerical results of Born cross sections are listed in Table~\ref{tab:sec}.

\begin{table}[h!]
\begin{center}
\caption{ Integrated luminosities ($\mathcal{L}_{\rm int}$), detection efficiencies ($\epsilon$), signal yields ($N^{\rm sig}$), ISR factors $(1+\delta^{\gamma})$, VP factors $(\frac{1}{\vert 1-\Pi\vert^2})$, and the obtained Born cross sections ($\sigma^{\rm Born}$) at different c.m.~energies ($\sqrt{s}$). The first uncertainties for Born cross sections are statistical and the second are systematic.}\label{tab:sec}
\vspace{0.2cm}
\begin{small}
\begin{tabular}{cr@{.}l c r@{~$\pm$~}l cccc}
\hline\hline
 $\sqrt{s}$~(GeV)   & \multicolumn{2}{c}{$\mathcal{L}_{\rm int}~(\rm pb^{-1})$} &$\epsilon(\%)$  &\multicolumn{2}{c}{$N^{\rm sig}$} & $(1+\delta^{\gamma})$   & $\frac{1}{\vert 1-\Pi\vert^2}$  & $\sigma^{\rm Born}$ (pb)   \\ \hline
 4.0076  &482&0     &$3.9 $             &634&28	  &4.5        &1.0435  &$8.1 \pm0.4\pm0.6 $            \\
 4.1285  &393&4                &$4.4 $  &408&23	  &  4.6    &1.0526    &$5.5   \pm 0.3 \pm0.5           $\\
 4.1574  &406&9               &$4.2 $ &398 & 22	  &  4.8    &1.0535    &$5.1   \pm 0.3 \pm0.5          $\\
 4.1780  &3194&5	               &$4.1 $ &2888 & 60  &  4.8     &1.0548  &$4.9   \pm 0.1 \pm0.5          $\\
 4.1890  & 523&9    &$4.2 $          &452 & 24	  & 4.8      &1.0560   &$4.6   \pm 0.2 \pm0.4           $\\
 4.1990  & 525&2   &$4.2 $        &462 & 26	  &4.9        &1.0568  &$4.6   \pm 0.3 \pm0.5           $\\
 4.2093  & 517&2    &$4.1 $        &467 & 24	  & 4.8      &1.0565   &$4.9   \pm 0.3 \pm0.5          $\\
 4.2188  & 513&4    &$4.3 $         &444 & 24	  & 4.9      &1.0565   &$4.5   \pm 0.2 \pm0.4          $\\
 4.2263  & 1056&4   &$3.9 $        &909 & 34	  &  4.9    &1.0548    &$4.8   \pm 0.2 \pm0.4         $\\
 4.2358   & 529&1   &$4.0 $        &427 & 23	  &  5.0    &1.0554    &$4.3   \pm 0.2 \pm0.4         $\\
 4.2439   & 536&3  &$4.3 $         &459 & 24	  &  5.0    &1.0552    &$4.4   \pm 0.2 \pm0.4           $\\
 4.2580   & 828&4  &$3.9 $        &670 & 30	  & 5.0      &1.0533   &$4.5   \pm 0.2 \pm0.4           $\\
 4.2668   & 529&7  &$3.8 $         &430 & 13	  & 5.0      &1.0531   &$4.5   \pm 0.1 \pm0.4           $\\
 4.2777   & 175&2  &$3.7 $         &131 & 14	  & 5.1      &1.0529   &$4.3   \pm 0.5 \pm0.5          $\\
 4.2879  &491&5                &$3.9 $ &421 & 23	  &  5.1    &1.0525    &$4.6   \pm 0.3 \pm0.4          $\\
 4.3121  &492&1               &$3.7 $ &366 & 22	  &   5.2  &1.0519     &$4.2   \pm 0.3 \pm0.5        $\\
 4.3374  &501&1                &$3.7 $ &390 & 22	  & 5.2      &1.0508   &$4.3   \pm 0.2 \pm0.5          $\\
 4.3583   & 543&9  &$3.7 $            &377 & 22	  &5.3       &1.0511   &$3.8   \pm 0.2 \pm0.3          $\\
 4.3774  &522&8                &$3.8 $ &406 & 22	  &  5.4    &1.0514    &$4.1   \pm 0.2 \pm0.3         $\\
 4.3965  &505&0                &$3.4 $ &255 & 18	  & 5.4      &1.0517   &$2.9   \pm 0.2 \pm0.3           $\\
 4.4156   & 1043&9  &$3.7 $        &716 & 30	  &  5.4    &1.0524    &$3.7   \pm 0.2 \pm0.3         $\\
 4.4362  &568&1                &$3.7 $ &365 & 21	  &  5.5    &1.0543    &$3.4   \pm 0.2 \pm0.4          $\\
 4.4671  &111&1  &$3.6 $             &80 & 10	  &  5.6    &1.0548    &$3.8   \pm 0.5 \pm0.4         $\\
 4.5995   & 586&9  &$3.1$           &259 & 18	  &6.1        &1.0547  &$2.5   \pm 0.2 \pm0.2           $\\ \hline\hline
 \end{tabular}
\end{small}
\end{center}
\end{table}

\subsection{Born cross section for intermediate states}
\label{sec:born}
\begin{figure}[h!]
\centering
\includegraphics[width=1.0\textwidth]{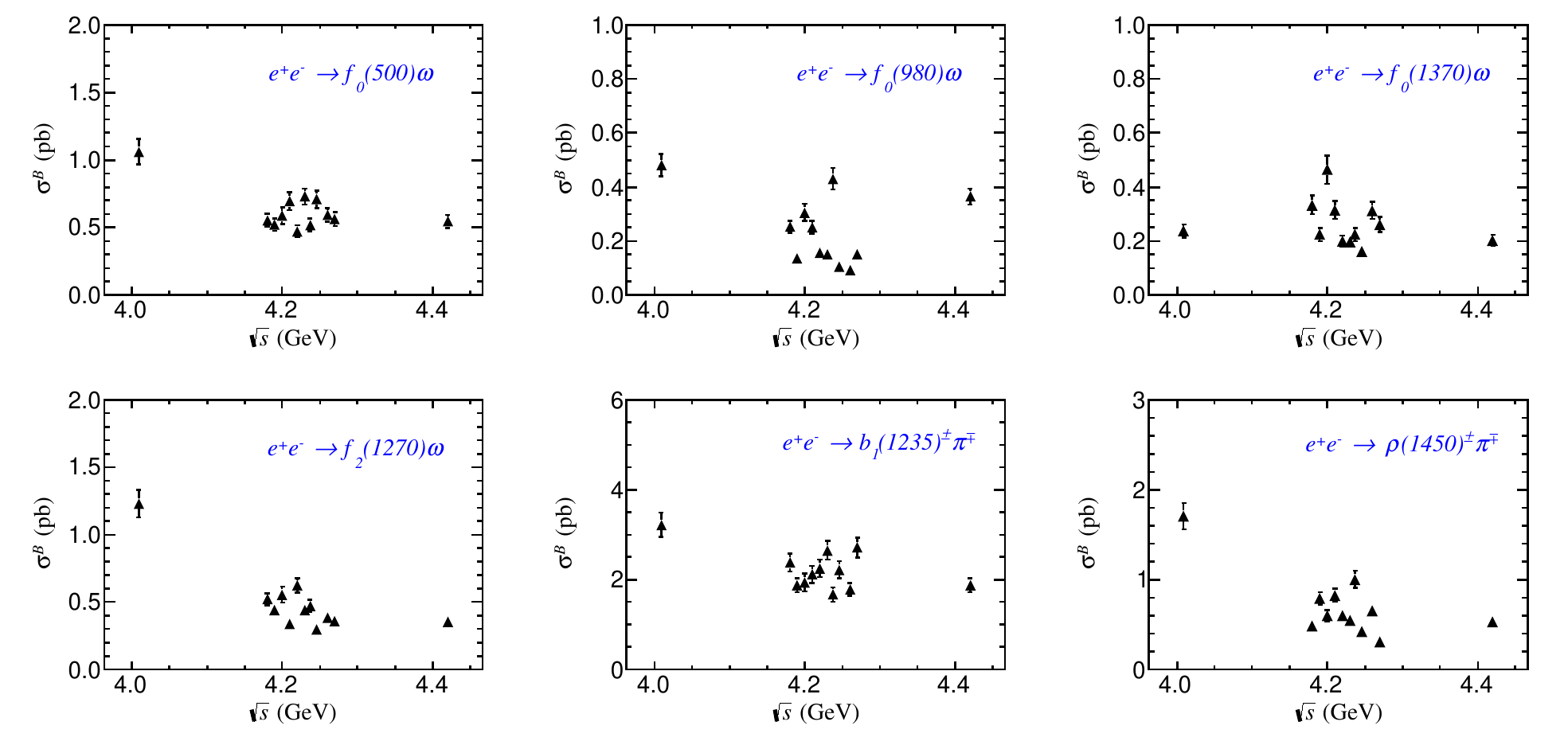}
\caption{ The Born cross sections of the processes containing $f_{0}(500)$, $f_{0}(980)$,
$f_{2}(1270)$, $f_{0}(1370)$, $b_{1}(1235)^{\pm}$, and $\rho(1450)^{\pm}$. Uncertainties combine both statistical and systematic uncertainties. }
\label{fig:sub}
\end{figure}

\hspace{1.5em}
The Born cross section for each intermediate state is calculated by
\begin{equation} \label{eq:modesc}
\sigma^{\rm Born}_{i} = R_i \, \sigma^{\rm Born} ,
\end{equation}
where $\sigma^{\rm Born}$ is the total Born cross section of $\ee \to \pip\pim\omega$, including the interference contributions among all intermediate states.
The cross-section ratio, $R_{i}$, is calculated according to Eq.~(\ref{yieldsFormula}) and given in Table~\ref{tabratio}, and the Born cross section of each intermediate state is shown in Fig.~\ref{fig:sub}.

\begin{table}[h!]
\centering{
\caption{The ratios of signal yields for intermediate states ($f_{0}(500)$, $f_{0}(980)$,
$f_{2}(1270)$, $f_{0}(1370)$, $b_{1}(1235)^{\pm}$, $\rho(1450)^{\pm}$), and non-resonant $\pip\pim\omega$ at different c.m.~energy points, which are  divided into groups A and B for higher statistics. }\label{tabratio}
\vspace{0.2cm}
\hspace*{-29pt}
\begin{small}
\setlength{\tabcolsep}{1.2mm}{
\begin{tabular}{clccccccc}
\hline\hline
Group &$\sqrt s$ (GeV) & $ f_{0}(500)$ &$ f_0(980)$ &$ f_0(1370)$ &$ f_2(1270)$  &$b_1(1235)^\pm$ &$\rho(1450)^{\pm}$ &$\pip\pim\omega$\\ \hline
 &4.0076&   $0.13\pm0.02$     &$0.06\pm0.02$       &$0.03 \pm0.02$       &$0.15 \pm0.03$      &$0.40 \pm0.05$       &$0.21\pm0.04$   &$0.00\pm0.01$   \\
 &4.1780&   $0.11\pm0.01$     &$0.05\pm0.01$       &$0.07\pm0.01$        &$0.11\pm0.01$       &$0.48\pm0.02$        &$0.10\pm0.02$     &$0.09\pm0.02$  \\
A&4.1890&   $0.11\pm0.02$     &$0.03\pm0.02$       &$0.05\pm0.03$        &$0.09\pm0.03$       &$0.40\pm0.05$        &$0.17\pm0.04$   &$0.12\pm0.04$    \\
 &4.1990&   $0.13\pm0.02$     &$0.07\pm0.03$       &$0.10\pm0.04$        &$0.12\pm0.04$       &$0.42\pm0.05$        &$0.13\pm0.04$      &$0.09\pm0.03$ \\
 &4.2093&   $0.14\pm0.03$     &$0.05\pm0.03$       &$0.06\pm0.04$        &$0.07\pm0.03$       &$0.43\pm0.05$        &$0.17\pm0.05$     &$0.06\pm0.03$  \\
 &4.2188&   $0.11\pm0.02$     &$0.03\pm0.02$       &$0.04\pm0.03$        &$0.14\pm0.03$       &$0.51\pm0.06$        &$0.13\pm0.05$    &$0.06\pm0.03$   \\ \hline

 &4.2263&   $0.15\pm0.02$     &$0.03\pm0.01$       &$0.04\pm0.02$        &$0.09\pm0.02$       &$0.56\pm0.04$        &$0.11\pm0.03$     &$0.03\pm0.03$  \\
 &4.2358&   $0.12\pm0.03$     &$0.10\pm0.03$       &$0.05\pm0.04$        &$0.11\pm0.02$       &$0.39\pm0.05$        &$0.24\pm0.03$     &$0.05\pm0.03$  \\
B&4.2439&   $0.16\pm0.03$     &$0.02\pm0.02$       &$0.04\pm0.02$        &$0.07\pm0.02$       &$0.51\pm0.05$        &$0.10\pm0.04$   &$0.10\pm0.04$    \\
 &4.2580&   $0.13\pm0.02$     &$0.02\pm0.01$       &$0.07\pm0.03$        &$0.09\pm0.02$       &$0.40\pm0.04$        &$0.15\pm0.03$     &$0.11\pm0.03$  \\
 &4.2668&   $0.12\pm0.02$     &$0.03\pm0.02$       &$0.06\pm0.04$        &$0.08\pm0.03$       &$0.60\pm0.07$        &$0.07\pm0.04$    &$0.03\pm0.03$   \\
 &4.4156&   $0.15\pm0.02$     &$0.10\pm0.03$       &$0.05\pm0.03$        &$0.10\pm0.02$       &$0.51\pm0.05$        &$0.14\pm0.04$    &$0.01\pm0.01$   \\ \hline\hline
\end{tabular}}
\end{small}}
\end{table}

\begin{table}[h!]
\centering{
\caption{The signal yields for intermediate states ($f_{0}(500)$, $f_{0}(980)$,
$f_{2}(1270)$, $f_{0}(1370)$, $b_{1}(1235)^{\pm}$, and $\rho(1450)^{\pm}$) at different c.m.~energy points, which are  divided into groups A and B for higher statistics. The errors are statistical only. }\label{tabyield}
\vspace{0.2cm}
\hspace*{-29pt}
\begin{small}
\setlength{\tabcolsep}{0.9mm}{
\begin{tabular}{cc r@{~$\pm$~}l r@{~$\pm$~}l r@{~$\pm$~}l r@{~$\pm$~}l r@{~$\pm$~}l r@{~$\pm$~}l}
\hline\hline
Group &$\sqrt s$ (GeV) & \multicolumn{2}{c}{$ f_{0}(500)$} & \multicolumn{2}{c}{$ f_0(980)$} & \multicolumn{2}{c}{$ f_0(1370)$} & \multicolumn{2}{c}{$ f_2(1270)$}  & \multicolumn{2}{c}{$b_1(1235)^\pm$} & \multicolumn{2}{c}{$\rho(1450)^{\pm}$}\\ \hline
 &4.0076 &77.40&13.51     &34.88&12.36       &17.21&12.15       &89.32&15.50      &233.49&26.19       &123.89&23.47      \\
 &4.1780 &298.92&28.95     &136.22&29.90       &179.80&40.13        &281.23&37.93       &1284.20&63.06        &260.12&42.96       \\
A&4.1890 &47.14&10.67     &12.28&11.92       &20.31&13.80       &39.90&13.47       &170.09&19.53        &71.91&17.73       \\
 &4.1990 &53.41&11.19     &27.73&14.27       &42.27&16.92        &50.58&15.52       &176.50&19.13        &54.62&16.57       \\
 &4.2093 &59.40&12.32     &21.34&11.31       &26.77&17.14        &28.84&13.23       &180.88&19.48        &70.47&19.60       \\
 &4.2188 &42.73&10.00     &13.99&9.97       &17.97&13.92        &56.45&13.82       &203.56&22.39        &54.25&19.80       \\ \hline
 &4.2263 &128.26&17.67     &26.69&12.74       &34.71&18.75        &77.60&17.80       &466.33&33.06        &95.40&24.50       \\
 &4.2358 & 47.61&12.54     &39.58&13.21       &20.56&15.28        &43.47&10.32       &153.61&20.45        &92.14&19.50       \\
B&4.2439 &68.69&11.56     &9.99&7.78       &15.42&11.29        &28.71&11.16       &214.59&20.95        &41.08&18.88       \\
 &4.2580 &81.67&15.42     &12.52&8.91       &43.09&16.44        &53.24&14.53       &246.22&24.71        &89.86&21.01       \\
 &4.2668 &49.34&11.15     &13.15&8.74       &22.76&16.21        &31.61&13.65       &237.85&26.53        &27.10&18.16       \\
 &4.4156 &96.94&16.25     &64.90&20.71       &35.83&26.22        &62.95&17.77       &334.32&28.10       &94.23&24.30       \\ \hline\hline
\end{tabular}}
\end{small}}
\end{table}

\section{SYSTEMATIC UNCERTAINTY}
\subsection{Uncertainty of the Born cross section}
\label{sec:systematic}
\hspace{1.5em}
The uncertainties in the Born cross section measurements arise from the luminosity
measurement, tracking and PID efficiency, photon detection efficiency, branching fraction, $K_S^0$ veto, ISR correction, fit procedure, PWA, and insignificant resonances.
However, the effects of the $\rm E_{EMC}/\it p$ requirement and $\chi_{c0}$ veto on efficiency are negligible.

\begin{itemize}
\item {\it Luminosity.} The integrated luminosity is measured by the Bhabha scattering process, and the uncertainty is $1.0$\%~\cite{rlum}.

\item {\it Tracking and PID efficiencies.} The uncertainty of the tracking efficiency has been studied with a high purity control sample of $e^+e^-\to \pi^+\pi^-K^+K^-$~\cite{prd1012003}.
The differences of the tracking and PID efficiencies between data and MC simulation in different transverse momentum and momentum ranges are taken as the systematic uncertainties of tracking and PID efficiencies, both $1.0$\% per charged pion.

\item {\it Photon detection efficiency.} The uncertainty from the photon detection has been studied
with the control samples of $\psi(3686)\to\pi^0\pi^0 J/\psi$ and $e^+e^-\to \omega\pi^0\to \pi^+\pi^-\pi^0\pi^0$~\cite{prd1012003}, which is $1.0$\% per photon.

\item {\it Branching fraction.} The branching fractions $Br(\omega\to\pi^+\pi^-\pi^0)$ and $Br(\pi^0\to\gamma\gamma)$ are quoted from the PDG~\cite{pdg}, which are
($89.2\pm0.7$)\% and ($98.823\pm0.034$)\%, respectively.
The relevant systematic uncertainty is $0.75$\% in total.

\item {\it $K_S^0$ veto.} The uncertainty of $K_S^0$ veto is taken as the difference of efficiencies with and without
$K_S^0$ veto between data and MC simulation, which is $0.8$\%.

\item {\it ISR correction.} To obtain reliable detection efficiencies, the Born cross sections input in the generator have been iterated until the $(1+\delta^{\mathit{r}})\cdot\epsilon$ values converge.
The differences of $(1+\delta^{\mathit{r}})\cdot\epsilon$ between the last two iterations are taken as the corresponding systematic uncertainties.

\item {\it Fit procedure.} The systematic uncertainty in the fit of $M_{\pip\pim\piz}$ mainly comes from the fit range,
signal shape and background shape. The fit range is changed from [$0.68$, $0.91$] GeV/$c^2$ to
[$0.67$, $0.92$] GeV/$c^2$.
The signal shape is changed to the $BW$ function convolved with a Gaussian resolution function.
The background shape is changed from the second-order Chebyshev polynomial to the third-order, and the
parameter of the background function is fixed to that derived from the fit to the largest data sample taken at $\sqrt s=4.178$~GeV.
The quadrature sum of the changes in the fitted signal yield is taken as the uncertainty.

\item {\it PWA.} The uncertainties due to the mass and width of the intermediate resonance state,
the background level, and the kinematic fit are considered in the systematic uncertainty of PWA.
The main contribution comes from $f_{0}(500)$, $f_{0}(980)$,
$f_{2}(1270)$, $b_{1}(1235)^{\pm}$, and $\rho(1450)^\pm$.
The total uncertainty is the sum of the following three detailed sources.

\item {\it Mass and width.}
The masses and widths of the intermediate resonance states in
this analysis are fixed on the PDG values~\cite{pdg}.
To estimate their systematic uncertainties,
we shift the mass and width of each intermediate resonance within one standard deviation.

\item {\it Background level.}
The background level is determined by the $\omega$ sideband events of the data sample. It is the same size as the number of events obtained in the $\omega$ signal region after the integration of the background function.
To estimate the systematic uncertainty of the background level,
we determine the deviation of the background level according to $\Delta n = \sqrt N$,
where $N$ is the estimated number of the background events in the $\omega$ signal region,
and change the background yield by ($N+\Delta n$).

\item {\it Kinematic fit.}
The uncertainty of the kinematic fit is
estimated by correcting the helix parameters of the charged
tracks to improve the consistency between data and
MC simulation~\cite{prd012002}. The difference in the detection
efficiencies of the TOY MC samples is regarded as the systematic uncertainty in PWA.

\item{\it Insignificant resonance.} An intermediate state with significance less than $5\sigma$, $\rho(1570)^\pm$, is removed in the normal solution.
The uncertainty is defined as
the difference between the detection efficiencies of the normal solution with and without the $\rho(1570)^\pm$ contribution.

\end{itemize}

The numerical values of these systematic uncertainties are summarized in Table~\ref{tab:uncertainty}. For the total uncertainty these contributions are added in quadrature.

\subsection{Uncertainty of the Born cross section for intermediate process}
\label{sec:systematic}
\hspace{1.5em}
The systematic uncertainty in the measurements of the Born cross sections for the intermediate processes
is the same as that of $e^+e^- \to \pi^{+}\pi^{-}\omega$.
Whereas, for the Born cross section measurement of intermediate state, the uncertainty in PWA depends on the ratio of each
intermediate state, $R_i$.
We mainly estimate the systematic uncertainty of the Born cross section of different intermediate processes for the twelve c.m.~energy points with higher statistics.
Their uncertainties are obtained by adding the individual contributions in quadrature and summarized in Table~\ref{ratio_sys}.

\begin{itemize}
\item {\it Mass and width.} For the uncertainties of the mass and
width of any of the intermediate resonance states,
we change its mass and width according to the PDG values within $\pm 1\sigma$.

\item {\it Background level.} We determine the deviation of the background level according to ${n}_{i}$,
and change the background yield to obtain the uncertainty of the background level.

\item {\it Kinematic fit.} We use the PHSP signal MC sample corrected by the helix parameters to re-perform PWA to estimate the uncertainty of the kinematic fit.

\item{\it Insignificant resonance.} The uncertainty due to one insignificant resonance was defined as
the difference between the ratio of the normal solution with and without the $\rho(1570)^\pm$ contribution.
\end{itemize}

For each source, the deviation from the nominal result is taken as the corresponding systematic uncertainty.

\begin{table}[h!]
\begin{center}
\caption{ Relative systematic uncertainties (in \%) in the cross section measurements include the luminosity(Lum),
the tracking efficiency (Trk), the PID, the photon detection (PD), the branching fraction (BF), the veto of $K_S^0$ ($K_S^V$), the ISR correction (ISR),
the signal shape (SS), the background shape (BS), the fit range (FR), PWA, and insignificant resonance (IR).
The sources with a superscript * are the common systematic uncertainties for
different c.m.~energies. }\label{tab:uncertainty}
\vspace{0.2cm}
\begin{small}
\begin{tabular}{ccccccccccccccc}\hline\hline
 $\sqrt s$~(GeV)     &$\rm Lum^*$    &$\rm Trk^*$    &$\rm PID^*$  &$\rm PD^*$ &$\rm BF^*$ &$K_S^V$$^*$ &ISR &SS &BS &FR &PWA &IR &Total \\  \hline
 4.0076  &1.0	&4.0	&4.0	 &2.0 &0.75   & 0.8 & 1.1  &0.8   &1.1   &0.6   & 0.3    & 3.1     & 7.2     \\
 4.1285  &1.0  &4.0  &4.0   &2.0 &0.75 & 0.8  & 2.8  &0.7   &0.7   &1.0   & 0.2    & 4.1      & 8.1                       \\
 4.1574  &1.0  &4.0  &4.0   &2.0 &0.75  & 0.8 & 1.5  &1.0   &1.5   &0.0   & 4.8    & 0.4      & 8.2      \\
 4.1780  &1.0	&4.0	&4.0	 &2.0 &0.75  & 0.8 & 5.2  &0.6   &1.1   &0.7   & 0.0     & 1.7     & 8.4    \\
 4.1890  &1.0	&4.0	&4.0	 &2.0 &0.75 & 0.8  & 0.7  &0.9   &0.7   &0.0   & 3.5     & 3.0      & 7.8     \\
 4.1990  &1.0	&4.0	&4.0	 &2.0 &0.75  & 0.8 & 5.5  &0.0   &1.5   &0.7   & 2.6     & 4.6      & 10.0      \\
 4.2093  &1.0	&4.0	&4.0	 &2.0 &0.75  & 0.8 & 3.9  &0.6   &0.9   &1.2   & 2.7     & 2.6      & 8.3    \\
 4.2188  &1.0	&4.0	&4.0	 &2.0 &0.75  & 0.8 & 1.0  &0.7   &0.9   &1.1   & 1.7     & 5.8     & 8.9    \\
 4.2263  &1.0	&4.0	&4.0	 &2.0 &0.75  & 0.8 & 0.6  &0.3   &1.3   &0.4   & 1.3     & 3.9      & 7.6    \\
 4.2358  &1.0	&4.0	&4.0	 &2.0 &0.75  & 0.8 & 3.6  &0.2   &0.9   &1.9   & 2.1     & 2.9    & 8.3    \\
 4.2439  &1.0	&4.0	&4.0	 &2.0 &0.75 & 0.8  & 1.5  &0.7   &1.1   &0.7   & 0.8     & 3.1     & 7.3    \\
 4.2580  &1.0	&4.0	&4.0	 &2.0 &0.75  & 0.8 & 0.8  &1.4   &1.5   &1.1   & 2.7     & 1.4     & 7.3    \\
 4.2668  &1.0	&4.0	&4.0	 &2.0 &0.75 & 0.8  & 1.1  &0.2   &1.2   &0.0   & 1.5     & 4.0     & 7.7    \\
 4.2777  &1.0	&4.0	&4.0	 &2.0 &0.75 & 0.8  & 4.7  &0.0   &1.5   &0.0   & 1.5     & 4.6      & 9.3    \\
 4.2879  &1.0  &4.0  &4.0   &2.0 &0.75 & 0.8  & 2.7  &0.5   &1.0   &0.5   & 0.2    & 1.0       & 6.9    \\
 4.3121  &1.0  &4.0  &4.0   &2.0 &0.75  & 0.8 & 5.9  &0.6   &1.6   &1.1   & 0.9    & 5.0      & 10.1   \\
 4.3374  &1.0  &4.0  &4.0   &2.0 &0.75  & 0.8 & 6.8  &1.0   &0.5   &0.5   & 3.0    & 1.1       & 9.8   \\
 4.3583  &1.0	&4.0	&4.0	 &2.0 &0.75 & 0.8  & 3.2  &1.1   &1.3   &1.6   & 1.1     & 2.5     & 7.8   \\
 4.3774  &1.0  &4.0  &4.0   &2.0 &0.75 & 0.8  & 0.0  &0.5   &1.0   &1.5   & 0.2    & 0.5      & 6.5     \\
 4.3965  &1.0  &4.0  &4.0   &2.0 &0.75 & 0.8  & 2.9  &0.4   &0.8   &0.0   & 2.0    & 1.2        & 7.3     \\
 4.4156  &1.0	&4.0	&4.0	 &2.0 &0.75 & 0.8 & 1.4  &0.3   &0.8   &0.9   & 1.7      & 3.2    & 7.4      \\
 4.4362  &1.0  &4.0  &4.0   &2.0 &0.75  & 0.8 &5.5  &0.8   &1.1   &0.8   & 3.3     & 0.1     & 9.0     \\
 4.4671  &1.0  &4.0  &4.0   &2.0 &0.75 & 0.8  & 3.8  &0.0   &1.3   &1.3   & 2.8    & 5.6      & 9.7     \\
 4.5995  &1.0	&4.0	&4.0	 &2.0 &0.75 & 0.8  & 1.6  &0.8   &1.1   &1.6   & 3.0     & 2.4      & 7.7      \\
\hline\hline
 \end{tabular}
\end{small}
\end{center}
\end{table}

\begin{table}[h!]
\centering{
\caption{ The systematic uncertainties (in \%) in the cross section measurements of the intermediate processes containing $f_{0}(500)$, $f_{0}(980)$,
$f_{2}(1270)$, $f_{0}(1370)$, $b_{1}(1235)^{\pm}$, and $\rho(1450)^{\pm}$.
The results of simultaneous to groups A and B, which are combined from twelve c.m.~energy points for higher statistics.}\label{ratio_sys}
\vspace{0.2cm}
\begin{tabular}{cccccccc}
\hline \hline
Group &$\sqrt s$ (GeV) & $ f_0(500)$ & $ f_0(980)$ & $ f_0(1370)$ &$ f_2(1270)$  & $b_1(1235)^\pm$ &$\rho(1450)^\pm$ \\ \hline
 &4.0076    &7.9      &7.1      &9.1     &7.2     &7.2     &7.1                                                   \\
 &4.1780    &8.5      &8.3      &10.3     &8.3     &8.3     &8.2                                                   \\
A&4.1890    &7.0      &6.5      &9.4     &6.5     &6.5     &6.4                                                    \\
 &4.1990    &8.7    &8.7    &10.1   &8.7    &8.7    &8.6                                                  \\
 &4.2093    &7.7      &7.5      &9.3     &7.5     &7.6     &7.5                                                    \\
 &4.2188    &7.2      &6.9      &9.0     &6.9     &6.9     &6.8                                                   \\ \hline
 &4.2263    &6.9      &6.6      &8.4     &6.6     &6.6     &6.5                                                    \\
 &4.2358    &7.9      &7.7      &9.6     &7.7     &7.7     &7.6                                                    \\
B&4.2439    &7.3      &6.6      &8.9     &6.6     &6.7     &6.6                                                    \\
 &4.2580    &7.3      &6.7      &9.3     &6.6     &6.7     &6.8                                                    \\
 &4.2668    &8.5      &7.5      &10.2    &7.5     &7.6     &7.9                                                  \\
 &4.4156    &7.6      &6.9      &9.2    &6.8     &6.9     &7.1                                               \\ \hline \hline
\end{tabular}
}
\end{table}

\section{FIT TO THE LINE SHAPE }
\begin{figure}[h!]
\centering
\includegraphics[width=0.8\textwidth]{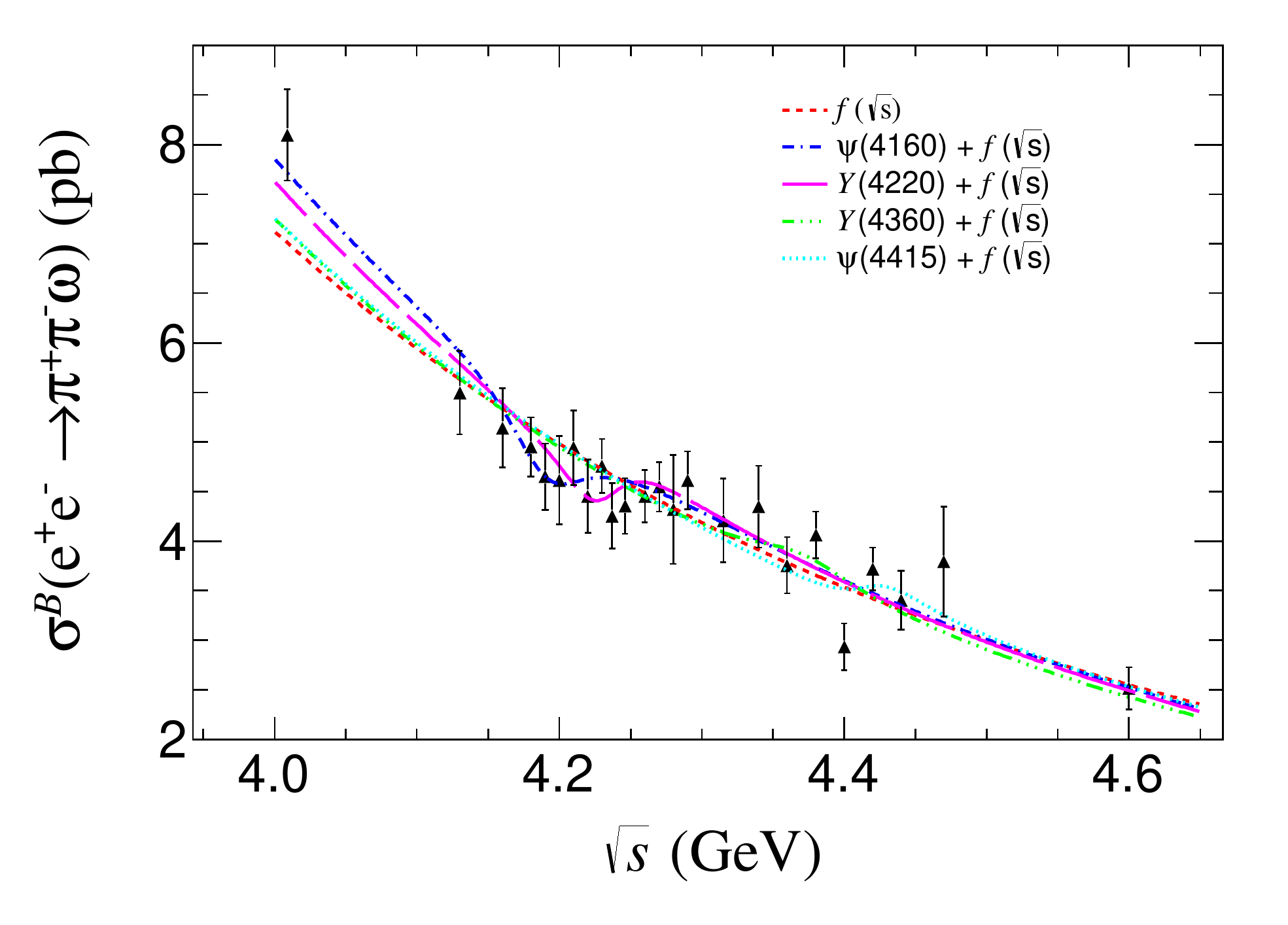}
\caption{ Fitted results of the measured Born cross sections at c.m.~energies between $4.0$ and $4.6$ GeV.
The data are presented as filled triangles with error bars combining
statistical and uncorrelated systematic uncertainties. The curves are the fit results to various amplitudes as  described in the text. }
\label{fig:fit1}
\end{figure}

\hspace{1.5em}
The line shape for total Born cross section of $\ee\to\pi^+\pi^-\omega$ is fitted with the
least square method~\cite{leastchi2}. First, the energy-dependent Born cross
section is parameterized by a non-resonant function $f(\sqrt{s}) = a/s^n$,
where $a$ and $n$ are free parameters. The correlations among different c.m.~energy
points are considered in the fit with the $\chi^2$ defined as below
(and  minimized by MINUIT~\cite{minuit}),
\begin{equation} \label{equ:chiq}
\mathcal{\chi}^2=\sum_{i}\frac{(\sigma_{B_{i}}- \sigma_{B_{i}}^{\rm fit})^2}{\delta^2_{i}} ,
 \end{equation}
where $\sigma_{B_{i}}$ and $\sigma_{B_{i}}^{\rm fit}$ are the measured and fitted values for Born
cross section at the $i$-th c.m.~energy point, respectively. here, $\delta_{i}$ is the uncertainty
for the $i$-th c.m.~energy point, which includes the statistical uncertainty and the uncorrelated part of the systematic uncertainty.
Figure~\ref{fig:fit1} shows the fit result with $\chi^2/\rm NDF = 27.75/(24-2-1) \approx 1.32$.

Secondly, the Born cross section is parameterized as the coherent sum of the energy-dependent
non-resonant function and one charmonium or charmonium-like state amplitude,
\begin{equation} \label{equ:eqbw}
\sigma^{\rm Born}(\sqrt{s}) = | BW( \sqrt{s})e^{i\phi} + \sqrt{\it f(\sqrt{s})} |^2 ,
\end{equation}
where $f(\sqrt{s})$ denotes the non-resonant amplitude, $\phi$ is the relative
phase between the continuum and resonant amplitudes, and $BW( \sqrt{s})$
is a relativistic $BW$ function which is used to describe the charmonium states,
$BW(\sqrt{s})=\frac{\sqrt{12\pi\Gamma_{ee}Br\Gamma_{\rm tot}}}{s-M^{2}+iM\Gamma_{\rm tot}}$.
And since these energies are far from the threshold of the $\ee\to\pi^+\pi^-\omega$ process, the effect of the three-body phase space factor is very small and therefore this $BW( \sqrt{s})$ function omits it.
The symbols $M$, $Br$, $\Gamma_{ee}$, and $\Gamma_{\rm tot}$ denote the mass, the branching
fraction of $\rm Y\to\pi^+\pi^-\omega$, the partial width to $\ee$, and the total width, respectively.
The considered charmonium and charmonium-like states
include $\psi(4160)$, $Y(4220)$, $Y(4360)$, and $\psi(4415)$.
In the fit, these resonance states are individually fitted with fixed mass and width from the PDG.
The fit results are shown in Fig.~\ref{fig:fit1}.
The goodness-of-fit tests for $\psi(4160)$, $Y (4220)$, $Y (4360)$, and  $\psi(4415)$ yield
$\chi^2/\rm NDF=$ $19.8/19$, $21.4/19$, $26.4/19$, and $26.6/19$, respectively.
The fit has two solutions with equal fit quality.
The fitted parameters of various resonance states are shown in Table~\ref{tab:par}.
The statistical significances of $\psi(4160)$ and $Y(4220)$ are $3.6\sigma$ and $3.1\sigma$,
while those of $Y(4360)$ and $\psi(4415)$ are $1.1\sigma$ and $1.0\sigma$, respectively.

\begin{table}[h!]
\begin{center}
\caption{ Fitted parameters and statistical significances for various charmonium states decaying into $\pi^+\pi^-\omega$. The uncertainties are statistical only.
 }\label{tab:par}
\begin{small}
\scalebox{0.8}{
\begin{tabular}{cccccccccc}\hline\hline
\multicolumn{1}{c}{\multirow{2}{1.5cm}{Parameter}} &\multicolumn{2}{c}{\textbf{$\psi(4160)$}} &\multicolumn{2}{c}{\textbf{$Y(4220)$}}   \\ \cline{2-5}
&\multicolumn{1}{c}{{Solution I}} &\multicolumn{1}{c}{{Solution II}} &\multicolumn{1}{c}{{Solution I}} &\multicolumn{1}{c}{{Solution II}} \\ \cline{1-5}
$12\pi\Gamma_{ee}Br$ (eV) &$0.03\pm0.02$  &$24.57\pm0.47$ &$18.29\pm0.32$ &$0.02\pm0.02$  \\
$\Gamma_{\rm tot}$ $(\rm GeV)$ &\multicolumn{2}{c}{0.070 } &\multicolumn{2}{c}{0.055}  \\
 M $(\rm GeV/c^{2})$   &\multicolumn{2}{c}{4.191} &\multicolumn{2}{c}{4.23} \\
$\phi$ (rad)   &$4.61\pm0.34$ &$4.68\pm0.01$ &$4.70\pm0.01$ &$5.38\pm0.31$  \\
 Significance ($\sigma$)    &\multicolumn{2}{c}{3.6} &\multicolumn{2}{c}{3.1}  \\
\hline\hline
 \end{tabular}
  }
\end{small}
\end{center}
\end{table}

\section{SUMMARY}
\label{sec:summary}
\hspace{1.5em}
In conclusion, the process of $\ee\to \pi^{+}\pi^{-}\omega$ is studied at twenty-four c.m.~energies
in the region from $4.0$ to $4.6$ GeV. The Born cross sections of $\ee\to \pi^{+}\pi^{-}\omega$
and the intermediate state production at twelve c.m.~energy points are measured with helicity amplitude analysis method.
The results indicate that the dominant contributions are from $\ee\to f_{0}(500)\omega$,
 $f_{0}(980)\omega$, $f_{2}(1270)\omega$, $f_{0}(1370)\omega$, $b_{1}(1235)^{\pm}\pi^{\mp}$,
 $\rho(1450)^{\pm}\pi^{\mp}$ with statistical significances greater than $5\sigma$.
By analyzing the line shape of the Born cross section of the $e^+e^-\to \pi^+\pi^-\omega$ process, greater than $3\sigma$ evidence for a state with mass about $4.2$~GeV/$c^2$ is found,
which is consistent with the production of either $\psi(4160)$ or $Y(4220)$.

\acknowledgments
\hspace{1.5em}
The BESIII collaboration thanks the staff of BEPCII and the IHEP computing center for their strong support. This work is supported in part by National Key R\&D Program of China under Contracts Nos. 2020YFA0406300, 2020YFA0406400; National Natural Science Foundation of China (NSFC) under Contracts Nos. 11975118, 12175244, 11875262, 11635010, 11735014, 11835012, 11935015, 11935016, 11935018, 11961141012, 12022510, 12025502, 12035009, 12035013, 12192260, 12192261, 12192262, 12192263, 12192264, 12192265, 12061131003; the Science and Technology Innovation Program of Hunan Province under Contract No. 2020RC3054; the Chinese Academy of Sciences (CAS) Large-Scale Scientific Facility Program; Joint Large-Scale Scientific Facility Funds of the NSFC and CAS under Contract No. U1832207; the CAS Center for Excellence in Particle Physics (CCEPP); 100 Talents Program of CAS; The Institute of Nuclear and Particle Physics (INPAC) and Shanghai Key Laboratory for Particle Physics and Cosmology; ERC under Contract No. 758462; European Union's Horizon 2020 research and innovation programme under Marie Sklodowska-Curie grant agreement under Contract No. 894790; German Research Foundation DFG under Contracts Nos. 443159800, Collaborative Research Center CRC 1044, GRK 2149; Istituto Nazionale di Fisica Nucleare, Italy; Ministry of Development of Turkey under Contract No. DPT2006K-120470; National Science and Technology fund; National Science Research and Innovation Fund (NSRF) via the Program Management Unit for Human Resources \& Institutional Development, Research and Innovation under Contract No. B16F640076; STFC (United Kingdom); Suranaree University of Technology (SUT), Thailand Science Research and Innovation (TSRI), and National Science Research and Innovation Fund (NSRF) under Contract No. 160355; The Royal Society, UK under Contracts Nos. DH140054, DH160214; The Swedish Research Council; U. S. Department of Energy under Contract No. DE-FG02-05ER41374

\bibliographystyle{IEEEtran}

\newpage
\begin{small}
\begin{center}
M.~Ablikim$^{1}$, M.~N.~Achasov$^{11,b}$, P.~Adlarson$^{70}$, M.~Albrecht$^{4}$, R.~Aliberti$^{31}$, A.~Amoroso$^{69A,69C}$, M.~R.~An$^{35}$, Q.~An$^{66,53}$, Y.~Bai$^{52}$, O.~Bakina$^{32}$
% ...
\\
\vspace{0.2cm}
(BESIII Collaboration)\\
\vspace{0.2cm} {\it
$^{1}$ Institute of High Energy Physics, Beijing 100049, People's Republic of China\\
% ...
$^{4}$ Bochum  Ruhr-University, D-44780 Bochum, Germany\\
% ...
$^{11}$ G.I. Budker Institute of Nuclear Physics SB RAS (BINP), Novosibirsk 630090, Russia\\
% ...
$^{31}$ Johannes Gutenberg University of Mainz, Johann-Joachim-Becher-Weg 45, D-55099 Mainz, Germany\\
$^{32}$ Joint Institute for Nuclear Research, 141980 Dubna, Moscow region, Russia\\
% ...
$^{35}$ Liaoning Normal University, Dalian 116029, People's Republic of China\\
% ...
$^{52}$ Southeast University, Nanjing 211100, People's Republic of China\\
$^{53}$ State Key Laboratory of Particle Detection and Electronics, Beijing 100049, Hefei 230026, People's Republic of China\\
% ...
$^{66}$ University of Science and Technology of China, Hefei 230026, People's Republic of China\\
% ...
$^{69}$ University of Turin and INFN, (A)University of Turin, I-10125, Turin, Italy; (B)University of Eastern Piedmont, I-15121, Alessandria, Italy; (C)INFN, I-10125, Turin, Italy\\
$^{70}$ Uppsala University, Box 516, SE-75120 Uppsala, Sweden\\
% ...
$^{b}$ Also at the Novosibirsk State University, Novosibirsk, 630090, Russia\\
% ...

}\end{center}

\end{small}

\end{document}